\documentclass[12pt]{article}

\usepackage{fullpage}
\usepackage{amssymb}
\usepackage{url}
\usepackage{bm}
\usepackage{color}

\usepackage{graphicx}

\newtheorem{theorem}{Theorem}
\newtheorem{lemma}{Lemma}
\newtheorem{corollary}{Corollary}

\newtheorem{property}{Property}

\newenvironment{proof}{\noindent Proof:}{$\square$}

\def\v#1{\mathbf{#1}}
\def\R{\mathbb{R}}

\def\proj{\mathrm{Proj}}

\def\thetav{{\bm\theta}}

\def\dx{\mathrm{d}x}
\def\dxx{\mathrm{d}\xx}

\def\muv{{\bm \mu}}

\def\Sigmam{{\bm\Sigma}}

\def\ballE{B}
\def\ball{B_F}
\def\balll{B_{F^*}}

\def\thetav{{\bm\theta}}

\def\vector#1{ \mathbf{#1} }
\def\set#1{\mathcal{#1}}

\def\alphav{{\bm\alpha}}

\newcommand{\pp}{\v{p}}
\newcommand{\qq}{\v{q}}
\newcommand{\xx}{\v{x}}
\newcommand{\yy}{\v{y}}
\newcommand{\ttt}{\v{t}}

\def\KL{\mathrm{KL}}

\def\gradF{{{\bm\nabla}F}}
\def\hessF{{{\bm\nabla ^2F}}}
\def\gradIF{{{\bm\nabla ^{-1}F}}}

\def\gradFstar{{{\bm\nabla}{F^*}}}
\def\dotproduct#1#2{{\langle {#1}, {#2} \rangle}}
\def\Dotproduct#1#2{{\left\langle {#1}, {#2} \right\rangle}}

\def\vector#1{\mathbf{#1}}
\def\set#1{\mathcal{#1}}
\def\argmin{\mathrm{argmin}}

\def\floor#1{\lfloor {#1} \rfloor}

\def\twovector#1#2{\left[\begin{array}{c}#1\\ #2\end{array}\right]}
\def\pow{\mathrm{pow}}

\def\vor{\mathrm{vor}}
\def\Vor{\mathrm{Vor}}
\def\del{\mathrm{del}}
\def\epi{\mathrm{epi}}
\def\reg{\mathrm{reg}}

\def\p{\mathbf{Pr}}
\def\twovector#1#2{\left[\begin{array}{c}#1\\ #2\end{array}\right]}
\def\muv{{\bm\mu}}

\def\dvx{\mathrm{d}\v{x}}
\def\grad{\bm\nabla}
\def\equaldef{ \stackrel{\mathrm{def}}{=} }

\def\v#1{\mathbf{#1}}
\def\R{\mathbb{R}}

\def\proj{\mathrm{Proj}}

\def\thetav{{\bm\theta}}

\def\dx{\mathrm{d}x}

\def\muv{{\bm \mu}}

\def\Sigmam{{\bm\Sigma}}

\def\thetav{{\bm\theta}}

\def\vector#1{ \mathbf{#1} }
\def\set#1{\mathcal{#1}}

\def\KL{\mathrm{KL}}

\def\dotproduct#1#2{{\langle {#1}, {#2} \rangle}}

\def\vector#1{\mathbf{#1}}
\def\set#1{\mathcal{#1}}
\def\argmin{\mathrm{argmin}}

\def\floor#1{\lfloor {#1} \rfloor}

\def\twovector#1#2{\left[\begin{array}{c}#1\\ #2\end{array}\right]}
\def\pow{\mathrm{pow}}

\def\vor{\mathrm{vor}}
\def\Vor{\mathrm{Vor}}

\def\p{\mathbf{Pr}}
\def\twovector#1#2{\left[\begin{array}{c}#1\\ #2\end{array}\right]}
\def\muv{{\bm\mu}}

\def\equaldef{ \stackrel{\mathrm{def}}{=} }

\begin{document}

\title{\Large Bregman Voronoi Diagrams: Properties, Algorithms and Applications\thanks{A preliminary version appeared in the 18th ACM-SIAM Symposium on Discrete Algorithms (SODA), pp.~746-755, 2007. Related materials are available online at \protect\url{http://www.csl.sony.co.jp/person/nielsen/BregmanVoronoi/} --- Version for arXiv (small resolution jpg/png files).}}
\author{Frank Nielsen\thanks{Sony Computer Science Laboratories Inc., Fundamental Research Laboratory, Japan.}
 \and Jean-Daniel Boissonnat\thanks{INRIA Sophia-Antipolis, GEOMETRICA, France.} 
 \and Richard Nock\thanks{Universit\'e Antilles-Guyane, CEREGMIA, France.}
}

\date{}

\maketitle

\begin{abstract} 
The Voronoi diagram of a finite  set of objects is a fundamental geometric
structure that subdivides the embedding space into regions, each
region consisting of the points that are closer to a given object than
to the others. We may define many variants of Voronoi diagrams
depending on the class of objects, the distance functions and the
embedding space.  In this paper, we investigate a framework for
defining and building Voronoi diagrams for a broad class of distance
functions called Bregman divergences. Bregman divergences include not
only the traditional (squared) Euclidean distance but also various
divergence measures based on entropic functions. Accordingly, Bregman
Voronoi diagrams allow to define information-theoretic Voronoi
diagrams in statistical parametric spaces based on the relative
entropy of distributions. We define several types of Bregman diagrams,
establish correspondences between those diagrams (using the
Legendre transformation), and show how to compute them
efficiently. We also introduce extensions of these diagrams,
e.g. $k$-order and $k$-bag Bregman Voronoi diagrams, and introduce Bregman
triangulations of a set of points and their connexion with  Bregman
Voronoi diagrams.  We show that these triangulations capture many of
the properties of the celebrated Delaunay triangulation. Finally, we give some
applications of Bregman Voronoi diagrams which are of interest in the
context of computational geometry and machine learning.

\end{abstract}

{\bf Categories and Subject Descriptors}: 
I.3.5 [{\bf Computer Graphics}] Computational Geometry and Object Modeling --- Geometric algorithms, languages, and systems;
F.2.2 [{\bf Analysis of Algorithms and Problem Complexity}]: Nonnumerical Algorithms and Problems --- Geometrical problems and computations;
G.2.1 [{\bf Discrete Mathematics}]: Combinatorics.

{\bf General Terms:} Algorithms, Theory


\section{Introduction and prior work}
The {\it Voronoi diagram} $\vor(\set{S})$ of a set of $n$ points
$\set{S}=\{\v{p}_1, ..., \v{p}_n\}$ of the $d$-dimensional Euclidean
space $\mathbb{R}^d$ is defined as the {\it cell complex} 
whose $d$-cells are the
{\it Voronoi regions} $\{\vor(\v{p}_i)\}_{i\in\{1, .., n\}}$ where
$\vor(\v{p}_i)$ is the set of points of
$\mathbb{R}^d$ closer to $\v{p}_i$ than to any other point of
$\set{S}$ with respect to a {\em distance function} $\delta$: 
$$\vor(\v{p}_i)\equaldef\{\v{x}\in\mathbb{R}^d\ |\
\delta (\v{p}_i,\v{x}) \leq \delta (\v{p}_j,\v{x}) \ \forall\
\v{p}_j\in\set{S} \}. $$ 
 Points $\{\vector{p}_i\}_i$ are called the {\it Voronoi sites} or
 {\it Voronoi generators}. Since its inception in disguise by
 Descartes in the 17th century~\cite{ak-vd-00}, Voronoi diagrams have
 found a broad spectrum of applications in science.  Computational
 geometers have focused at first on {\it Euclidean} Voronoi
 diagrams~\cite{ak-vd-00} by considering the case where $\delta (\xx
 ,\yy)$ is the Euclidean distance
 $||\xx-\yy||=\sqrt{\sum_{i=1}^d (x_i-y_i)^2}$. Voronoi diagrams have
 been later on defined and studied for other distance functions, most notably
 the $L_1$ distance $||\xx-\yy||_1=\sum_{i=1}^d |x_i-y_i|$ (Manhattan
 distance) and the $L_\infty$ distance $||\xx-\yy ||_\infty=
 \max_{i\in\{1, ..., d\}} |x_i-y_i|$~\cite{compgeom-1998,ak-vd-00}.  Klein
 further presented an {\em abstract framework} for describing and
 computing the fundamental structures of abstract Voronoi
 diagrams~\cite{abstractvoronoidiagrams-1989,bwy-cvd-07}.

\begin{figure}
\centering
\includegraphics[bb=0 0 180 205 , width=7cm]{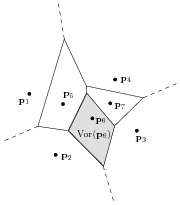}
\caption{Ordinary Euclidean Voronoi diagram of a given set $\set{S}$ of seven sites. 
In the bounded Voronoi cell $\vor(\v{p}_6)$, every point $\v{p}\in\vor(\v{p}_6)$ is closer to $\v{p}_6$ than to any other site of $\set{S}$
(with respect to the Euclidean distance).
Dashed segments denote infinite edges delimiting unbounded cells.  \label{fig:ordinaryl22}}
\end{figure}

In artificial intelligence, machine learning techniques also rely on
geometric concepts for {\it building classifiers} in supervised
problems ({\it e.g.}, linear separators, oblique decision trees, etc.) 
or {\it clustering data} in unsupervised settings ({\it e.g.},
$k$-means, support vector clustering~\cite{svc}, etc.).  However, the considered
data sets $\mathcal{S}$ and their underlying spaces $\set{X}$ are
usually {\it not} metric spaces.  The notion of distance between two
elements of $\set{X}$ needs to be replaced by a {\it pseudo-distance}
that is not necessarily symmetric and may not satisfy the {\it
triangle inequality}. Such a pseudo-distance is also referred to as
{\it distortion}, {\it (dis)similarity} or {\it divergence} in the
literature. For example, in parametric statistical spaces $\set{X}$, a
vector point represent a distribution and its coordinates store the
parameters of the associated distribution. A notion of ``distance''
between two such points is then needed to represent the divergence
between the corresponding distributions.

Very few works have tackled an in-depth study of Voronoi diagrams and
their applications for such a kind of statistical spaces. This is all
the more important even for ordinary Voronoi diagrams as Euclidean
point location of sites are usually {\em observed} in {\it noisy}
environments ({\it e.g.}, imprecise point measures in computer vision
experiments), and ``noise'' is often modeled by means of Normal
distributions (so-called ``Gaussian noise'').
To the best of our knowledge, statistical Voronoi diagrams have only
been considered in a 4-page short paper of Onishi and
Imai~\cite{vdnormal} which relies on Kullback-Leibler divergence of
$d$D multivariate normal distributions to study combinatorics of their
Voronoi diagrams, and subsequently in a 2-page video paper of Sadakane
et al.~\cite{voronoidivergence-1998} which defines the divergence
implied by a convex function and its conjugate, and present the
Voronoi diagram with flavors of information
geometry~\cite{informationgeometry} (see also~\cite{VoronoiExpFamily-1997} and related short communications~\cite{DBLP:conf/cccg/InabaI00,cccg98-inaba-geometric}). Our study of Bregman Voronoi
diagrams generalizes and subsumes these preliminary studies using an easier
concept of divergence: Bregman divergences~\cite{Bregman67,j-cbd-2005}
that do not rely {\it explicitly} on convex conjugates.
Bregman divergences encapsulate the squared Euclidean distance and many widely used divergences, e.g. the Kullback-Leibler divergence.
It should be noticed however that other divergences have been defined and studied in the context of Riemannian geometry~\cite{informationgeometry}.
Sacrifying for some generality, while not very restrictive in practice, allows a much simpler treatment and our study of Bregman divergences is elementary and does not rely on Riemannian geometry.

In this paper, we give a thorough treatment of Bregman Voronoi
diagrams which elegantly {\it unifies} the ordinary Euclidean Voronoi
diagram and statistical Voronoi diagrams.  Our contributions are
summarized as follows:

\begin{itemize}
\item Since Bregman divergences are not symmetric, we define {\em two types} 
of Bregman Voronoi diagrams. One is an affine diagram with convex
polyhedral cells while the other one is curved. The cells of those two
diagrams are in 1-1 correspondence through the Legendre
transformation.  We also introduce a {\em third-type} symmetrized
Bregman Voronoi diagram.

\item We present a simple way to compute the Bregman Voronoi diagram
of a set of points by lifting the points in a higher dimensional space
using an extra dimension. This mapping leads also to combinatorial
bounds on the size of these diagrams. We also define weighted Bregman
Voronoi diagrams and show that the class of these diagrams is
identical to the class of affine (or power) diagrams.  Special cases
of weighted Bregman Voronoi diagrams are the $k$-order and $k$-bag Bregman Voronoi
diagrams.

\item We define two  triangulations of a set of points. 
The first one captures some of the most important properties of the
well-known Delaunay triangulation.  The second triangulation is called
a geodesic Bregman triangulation since its edges are geodesic
arcs. Differently from the first triangulation, this triangulation is
the geometric dual of the first-type Bregman Voronoi diagram of its
vertices.

\item We give a few  applications of Bregman Voronoi diagrams which are of 
interest in the context of 
computational geometry and machine learning.

\end{itemize}

The outline of the paper is as follows: In Section~\ref{sec:Bregman},
we define Bregman divergences and recall some of their basic
properties.  In Section~\ref{sec-bregman-geometry}, we study the
geometry of Bregman spaces and characterize bisectors, balls and
geodesics. Section~\ref{sec:BVD} is devoted to Bregman Voronoi
diagrams and Section~\ref{sec:BregmanTriangulation} to Bregman
triangulations. In Section~\ref{sec:applications}, we select of few
applications of interest in computational geometry and machine learning.
Finally, Section~\ref{sec:Conclusion} concludes the paper and mention
further ongoing investigations.

\paragraph{Notations.} In the whole paper, $\set{X}$ denotes 
an open convex domain of $\mathbb{R}^d$ and $F: \set{X}
\mapsto \R$ a strictly convex and differentiable function.
$\mathcal{F}$ denotes the graph of $F$, i.e. the set of points $(\xx
,z)\in \set{X}\times \R$ where $z=F(\xx)$. We write $\hat{\xx}$ for the
point $(\xx,F(\xx))\in \mathcal{F}$. $\gradF$, $\hessF$ and $\gradIF$ denote respectively the gradient, the Hessian and the inverse gradient of $F$.

\section{Bregman divergences\label{sec:Bregman}}

In this section, we recall the definition of Bregman\footnote{Lev
M. Bregman historically pioneered this notion in the seminal
work~\cite{Bregman67} on minimization of a convex objective function
under linear constraints. See
\url{http://www.math.bgu.ac.il/serv/segel/bregman.html}. We gratefully acknowledge him for sending us this historical paper. } divergences and
some of their main properties (\S\ref{sec:Bregman:Definition}). We show
that the notion of Bregman divergence encapsulates the squared
Euclidean distance as well as several well-known information-theoretic divergences. We introduce the notion of
dual divergences (\S\ref{sec:Bregman:DualDivergences}) and show how
this comes in handy for symmetrizing Bregman divergences
(\S\ref{sec:Bregman:SymmetrizedDivergence}). 
Finally, we prove that the Kullback-Leibler divergence of distributions that belong to the exponential
family of distributions can be viewed as a Bregman divergence
(\S\ref{sec:Bregman:ExponentialFamilies}).

\subsection{Definition and basic properties\label{sec:Bregman:Definition}}

For any two points $\v{p}$ and $\v{q}$ of $\set{X}\subseteq\mathbb{R}^d$, the Bregman
divergence\footnote{See Java\texttrademark{} applet at \url{http://www.csl.sony.co.jp/person/nielsen/BregmanDivergence/}} $D_F(\cdot || \cdot): \set{X} \mapsto \mathbb{R}$ of $\v{p}$ to $\v{q}$ associated to a strictly convex and differentiable function $F$ (called the {\em
generator function} of the divergence) is defined as

\begin{equation}\label{eq:basicdf}
D_F(\v{p} || \v{q})  \equaldef  F(\v{p}) - F(\v{q}) - \dotproduct{\gradF (\v{q})}{\v{p} - \v{q}},
\end{equation}

\noindent where $\gradF=[\frac{\partial F}{\partial x_1}\ ...\  \frac{\partial F}{\partial x_d}]^T$ denotes the gradient operator, and $\dotproduct{\v{p}}{\v{q}}$ the inner (or dot) product: $\sum_{i=1}^d p_iq_i$.

Informally speaking, Bregman divergence $D_F$ is the {\it tail} of the
Taylor expansion of $F$. See~\cite{Csiszar91} for an axiomatic
characterization of Bregman divergences as ``permissible'' divergences.

\begin{lemma}\label{lem-code-BD}
The Bregman divergence $D_F(\v{p}||\v{q})$ is geometrically
measured as the vertical distance between $\hat{\v{p}}$ and the
hyperplane $H_{\v{q}}$ tangent to $\set{F}$ 
at point $\hat{\v{q}}$:  $D_F(\v{p}||\v{q})=F(\v{p})-H_{\v{q}}(\v{p})$.
\end{lemma}

\begin{figure}
\centering
\includegraphics[bb=0 0 431 340, width=7cm]{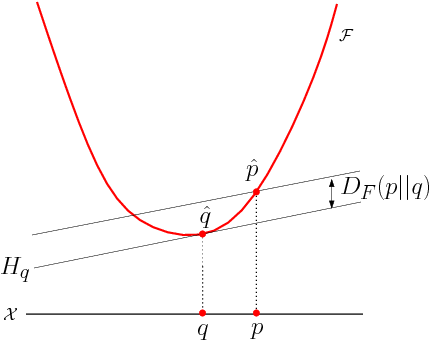}
\caption{Visualizing the Bregman divergence.
$D_F(.|| \v{q})$ is the
vertical distance between $\mathcal{F}$ and the hyperplane
tangent to  $\mathcal{F}$ at $\hat{\v{q}}$.
\label{fig:bregmanl22}}
\end{figure}

\begin{proof}
The tangent hyperplane to hypersurface $\set{F}: z=F(\v{x})$ at point $\hat{\v{q}}$ is $H_{\v{q}}: z= F(\v{q}) + \dotproduct{\gradF(\v{q})}{\v{x} - \v{q}}$.
It follows that $D_F(\v{p}||\v{q})=F(\v{p})-H_{\v{q}}(\v{p})$ (see Figure~\ref{fig:bregmanl22}).
\end{proof}

We now give some basic properties of Bregman divergences. 
The first property seems to be new. The others are well known.
First, observe that, for most functions $F$, the associated Bregman divergence is \textit{not}
symmetric, i.e. $D_F(\v{p}||
\v{q})\not =D_F(\v{q}||\v{p})$ (the symbol $||$ is put to emphasize
this point, as is standard in information theory). The following lemma proves this claim.

\begin{lemma}\label{notsym}
Let $F$ be properly defined for $D_F$ to exist. Then $D_F$ is symmetric if and only if  the Hessian $\hessF$  is constant on ${\set{X}}$.
\end{lemma}

\begin{proof}
($\Rightarrow$) From Eq.~\ref{eq:basicdf}, the symmetry  $D_F(\v{p}||\v{q})=D_F(\v{q}||\v{p})$ yields:
\begin{eqnarray}
F(\v{p}) & = & F(\v{q}) + \frac{1}{2}\langle \v{p} - \v{q}, \gradF(\v{q}) + \gradF(\v{p})\rangle\:\:. \label{csym}
\end{eqnarray}
A Taylor expansion of $F$ around $\v{q}$ using the Lagrange form of the remainder also yields:
\begin{eqnarray}
F(\v{p}) & = & F(\v{q}) + \langle \v{p} - \v{q}, \gradF(\v{q}) \rangle + \frac{1}{2} (\v{p} - \v{q})^T \hessF(\v{q}) (\v{p} - \v{q}) + \frac{1}{6} \langle \v{p} - \v{q}, \gradF\rangle^{3}(\v{r}_{\v{p}\v{q}})\:\:, \label{tex}
\end{eqnarray}
with $\v{r}_{\v{p}\v{q}}$ on the line segment $\v{p}\v{q}$. Equations (\ref{csym}) and (\ref{tex}) yield the following constraint:
\begin{eqnarray}
\langle \v{p} - \v{q}, \gradF(\v{p})\rangle & = & \langle \v{p} - \v{q}, \gradF(\v{q}) \rangle + (\v{p} - \v{q})^T \hessF(\v{q}) (\v{p} - \v{q}) + \frac{1}{3} \langle \v{p} - \v{q}, \gradF\rangle^{3}(\v{r}_{\v{p}\v{q}}) \:\:.\label{eq7}
\end{eqnarray}
On the other hand, if we make the Taylor expansion of $\gradF$ around $\v{q}$ and then multiply both sides by $\v{p} - \v{q}$, we separately obtain:
\begin{eqnarray*}
\langle \v{p} - \v{q}, \gradF(\v{p})\rangle & = & \langle \v{p} - \v{q}, \gradF(\v{q}) \rangle + (\v{p} - \v{q})^T \hessF(\v{q}) (\v{p} - \v{q}) + \frac{1}{2} \langle \v{p} - \v{q}, \gradF\rangle^{3}(\v{s}_{\v{p}\v{q}})\:\:,
\end{eqnarray*}
with $\v{s}_{\v{p}\v{q}}$ on the line segment $\v{p}\v{q}$. However, for this to equal Eq. (\ref{eq7}), we must have $\langle \v{p} - \v{q}, \gradF\rangle^{3}(\v{r}_{\v{p}\v{q}}) = (3/2) \langle \v{p} - \v{q}, \gradF\rangle^{3}(\v{s}_{\v{p}\v{q}})$ for each $\v{p}$ and $\v{q}$ in ${\set{X}}$. If we pick $\v{p}$ and $\v{q}$ very close to each other, this equality cannot be true, except when the third differentials are all zero on $\v{r}_{\v{p}\v{q}}$ and $\v{s}_{\v{p}\v{q}}$. Repeating this argument over each subset of $\set{X}$ having non zero measure, we obtain that the third differentials of $F$ must be zero everywhere but on subsets of $\set{X}$ with zero measure, which implies that the second differentials (the Hessian of $F$, $\hessF$) are {\em constant} everywhere on $\set{X}$.

($\Leftarrow$) Assume the hessian $\hessF$ is constant on $\set{X}$. In this case, because $F$ is strictly convex, the Hessian $\hessF$ is positive definite, and we can factor it as $\hessF = \v{P}^{-1} \v{D} \v{P}$ where $\v{D}$ is a diagonal matrix  and $\v{P}$ a unitary rotation matrix. Reasoning in the basis of ${\set{X}}$ formed by $\v{P}$, each element $\v{x}$ is mapped to $\v{P}\v{x}$, and we have $F(\v{x})=\sum_{i} {d_i x^2_i}$, where the $d_i$'s are the diagonal coefficients of $\v{D}$. The symmetry of $D_F$ is then immediate (i.e., $D_F$ is a generalized quadratic distance).
\end{proof}

\begin{property}[Non-negativity]
The strict convexity of generator function $F$ implies that, for any $\v{p}$ and $\v{q}$
in $\set{X}$, $D_F(\v{p}||\v{q})\geq 0$, with $D_F(\v{p}||\v{q})=0$ if
and only if $\v{p}=\v{q}$.
\end{property}

\begin{property}[Convexity]
Function $D_F(\v{p}||\v{q})$ is convex in its first argument $\v{p}$ but not 
necessarily in its second argument $\v{q}$.
\end{property}

Bregman divergences can easily be constructed from simpler ones.  For
instance, multivariate Bregman divergences $D_F$ can be created from
univariate generator functions coordinate-wise as $F(\v{x})=\sum_{i=1}^d
f_i(x_i)$ with $\gradF=[\frac{\mathrm{d} f_1}{\dx_1}\ ...\
\frac{\mathrm{d} f_d}{\dx_d}]^T$.

Because positive linear combinations of strictly  convex and differentiable
functions are strictly convex and differentiable functions, new generator functions
(and  corresponding Bregman divergences) can also be built as
positive linear combinations of elementary generator functions. This is an
important property as it allows to handle mixed data sets of
heterogenous types in a unified framework.

\begin{property}[Linearity] \label{prop:linearoperator}
 Bregman divergence is a linear operator, i.e., for any two strictly convex and differentiable functions $F_1$ and $F_2$ defined on $\set{X}$ and  for any $\lambda\geq 0$:
\[D_{F_1+\lambda F_2}(\v{p}||\v{q})=D_{F_1}(\v{p}||\v{q})+\lambda D_{F_2}(\v{p}||\v{q}).\]
\end{property}

\begin{property}[Invariance under linear transforms]\label{prop:invariance}
$G(\v{x})=F(\v{x})+\dotproduct{\v{a}}{\v{x}}+b$, with
$\v{a}\in\mathbb{R}^d$ and $b\in\mathbb{R}$, is a strictly convex and
differentiable function on $\set{X}$, and
$D_G(\v{p}||\v{q})=D_F(\v{p}||\v{q})$.
\end{property}

Examples of Bregman divergences are the squared Euclidean distance
(obtained for $F(\v{x})= \| \xx\| ^2$ and the generalized quadratic distance
function $F(\v{x})=\v{x}^T\v{Q}\v{x}$ where $\v{Q}$ is a positive
definite matrix.  When $\v{Q}$ is taken to be the inverse of the
variance-covariance matrix, $D_F$ is the Mahalanobis distance, extensively used
in computer vision.  More importantly, the notion of Bregman
divergence encapsulates various information measures based on entropic
functions such as the Kullback-Leibler divergence based on the
(unnormalized) Shannon entropy, or the Itakura-Saito divergence based
on Burg entropy (commonly used in sound
processing). Table~\ref{tab:fbreg} lists the main univariate Bregman divergences.

\begin{table}
\begin{center}
{\small
\begin{tabular}{|l||l||l|l||l|} \hline
Dom. $\set{X}$ & Function $F$ & Gradient   & Inv. grad.    & Divergence $D_F(p||q)$ \\ 
\hline\hline
$\mathbb{R}$ & Squared function  & & & Squared loss (norm) \\
 & $x^2$ & $2x$ & $\frac{x}{2}$ & $(p-q)^2$ \\ \hline 
  ${\mathbb R}_+, \alpha\in\mathbb{N}$ & Norm-like & & & Norm-like \\
   $\alpha>1$ & $x^\alpha$ & $\alpha x^{\alpha-1}$ & $(\frac{x}{\alpha})^{\frac{1}{\alpha-1}}$ & $p^\alpha+(\alpha-1)q^\alpha-\alpha pq^{\alpha-1}$ \\ \hline
   \hline
 $\mathbb{R}^+$ &  Unnorm. Shannon entropy & & & Kullback-Leibler div. (I-div.) \\
 & $x\log x-x$ & $\log x$ & $\exp(x)$ & $p\log \frac{p}{q}-p+q$ \\ \hline
 & Exponential & & & Exponential loss  \\  
$\mathbb{R}$ & $\exp x$ & $\exp x$ & $\log x$ &  $\exp(p)-(p-q+1)\exp(q)$ \\ \hline \hline
  $\mathbb{R}^+*$ & Burg entropy & & & Itakura-Saito divergence \\ 
   & $-\log x$  & $-\frac{1}{x}$ & $-\frac{1}{x}$ & $\frac{p}{q}-\log\frac{p}{q}-1$ \\ \hline \hline
   $[0,1]$ & Bit entropy &  &  & Logistic loss\\
 & $x\log x+(1-x)\log (1-x)$ & $\log\frac{x}{1-x}$ & $\frac{\exp x}{1+\exp x}$ & $p\log\frac{p}{q}+(1-p)\log\frac{1-p}{1-q}$ \\ \hline
& Dual bit entropy & & & Dual logistic loss \\
$\mathbb{R}$ &  $\log (1+\exp x)$ & $\frac{\exp x}{1+\exp x}$ & $\log \frac{x}{1-x}$ &  $\log\frac{1+\exp p}{1+\exp q}-(p-q)\frac{\exp q}{1+\exp q}$ \\ \hline \hline
$[-1,1]$  & Hellinger-like & & &  Hellinger-like \\
   & $-\sqrt{1-x^2}$ & $\frac{x}{\sqrt{1-x^2}}$ & $\frac{x}{\sqrt{1+x^2}}$ & $\frac{1-pq}{\sqrt{1-q^2}}-\sqrt{1-p^2}$ \\ \hline
   \hline \hline
   
\end{tabular}
}
\end{center}

\caption{Some common univariate Bregman divergences $D_F$. 
\label{tab:fbreg}}
\end{table}

\subsection{Legendre duality \label{sec:Bregman:DualDivergences}}

We now turn to an essential notion of convex analysis:
Legendre transform that will allow us to associate to any Bregman divergence a dual Bregman divergence.
  
Let $F$ be a strictly convex and differentiable real-valued function
on $\set{X}$. The Legendre
transformation makes use of the duality relationship between points and lines to associate to $F$ a  {\em convex conjugate}
function $F^*:\R^d\mapsto \R$ given by~\cite{ConvexAnalysis-1970}:

\[
F^*(\v{y})=\sup_{\v{x}\in\set{X}}  \{\dotproduct{\v{y}}{\v{x}} - F(\v{x}) \}.
\]

The supremum is reached at the {\it unique} point where the gradient
of $G(\v{x})=\dotproduct{\v{y}}{\v{x}} - F(\v{x})$ vanishes or,
equivalently, when $\v{y}=\gradF(\v{x})$.  

As is well-known, $F^*$ is strictly convex. To see this, consider the
epigraph $\epi (F^*)$, i.e. the set of points $(\yy,z)$ such that
$F^*(\yy ) \leq z$. Clearly, $(\yy ,z)\in \epi (F^*)$ iff
$G_{\xx}(\yy)=\dotproduct{\v{y}}{\v{x}} - F(\v{x}) \leq z$ for all $\xx\in
\set{X}$. Therefore, $\epi (F^*) =
\cap_{\v{x}\in\set{X}} \epi (G_{\xx})$. Since $G_{\xx}(\yy)$ is an affine function,
 $\epi (G_{\xx})$ is a half-space and $\epi (F^*)$ being the intersection of half-spaces is
a convex set, which proves that $F^*$ is convex. The strict convexity follows from the fact
that otherwise, $F$ would not be differentiable in at least one point $\v{z} \in \set{X}$: at this point,
$\dotproduct{\v{y}_\alpha}{\v{z}} - F(\v{z}) \geq \dotproduct{\v{y}_\alpha}{\v{x}} - F(\v{x}), \forall
\v{x} \in \set{X}$, and $\v{y}_\alpha = \alpha\v{y}_1 + (1-\alpha)\v{y}_2, \forall \alpha \in [0,1]$,
$\v{y}_1\v{y}_2$ being a segment on which $F^*$ is not strictly convex. Thus, $\v{y}_1\v{y}_2$ would
be a subdifferential of $F$ in $\v{z}$ contradicting the fact that $F$ is differentiable.

\begin{figure}
\centering
\includegraphics[bb=0 0 443 241 , width=11cm]{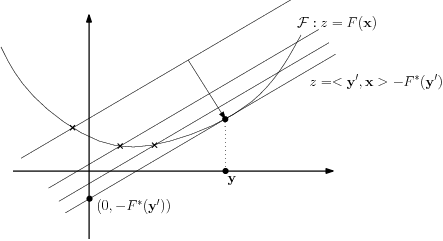}
\caption{Legendre transformation of a strictly convex function $F$: The $z$-intercept $(0,-F^*(\v{y}'))$ of the tangent hyperplane $H_{\v{y}} : z=\dotproduct{\v{y}'}{\v{x}}-F^*(\v{y}')$ of $\set{F}$ at $\hat{\v{y}}$ defines the value of the Legendre transform $F^*$ for the dual coordinate $\mathbf{y}'=\gradF(\v{y})$. Any hyperplane passing through an other point of $\set{F}$ and parallel to $H_{\v{y}}$ necessarily intersects the $z$-axis above $-F^*(\v{y}')$. \label{fig:legendre}}
\end{figure}

For convenience, we write $\v{x}'=\gradF(\v{x})$ (omitting the $F$ in
the $\v{x}'$ notation as it should be clear from the context).
Figure~\ref{fig:legendre} gives a geometric interpretation of the Legendre transformation.
Using this notation, Eq.~\ref{eq:basicdf} can be rewritten as 

\begin{equation}
D_F(\v{p}||\v{q})=F(\v{p})-F(\v{q})-\dotproduct{\v{q}'}{\v{p}-\v{q}}.
\label{eq-div'}
\end{equation}

Since $F$ is a strictly convex and differentiable real-valued function
on $\set{X}$, its gradient $\gradF$ is well defined as well as its
inverse $\gradIF$.  Writing $\set{X}'$ for the {\em gradient space} $\{
\gradF (\v{x})=\v{x}'|
\v{x}\in\set{X}\}$,  the convex conjugate 
$F^*$ of $F$ is the function: $\set{X}'\subset \R^d \mapsto \R$ defined by

\begin{equation}
F^*(\v{x}')=\dotproduct{\v{x}}{\v{x}'} - F(\v{x}).
\label{eq-conjugate}
\end{equation}

Deriving this expression, we get
\[ \dotproduct{\gradFstar (\xx ')}{\dxx '}= \dotproduct{\xx}{\dxx '}
+\dotproduct{\xx '}{\dxx} - \dotproduct{\gradF (\xx)}{\dxx} =
\dotproduct{\xx}{\dxx '} = \dotproduct{\gradIF(\xx ')}{\dxx '},\] from
which we deduce that $\gradFstar=\gradIF$. From
Eq.~\ref{eq-conjugate}, we also deduce $(F^*)^*=F$. 

From the above discussion, it follows that $D_{F^*}$ is a Bregman divergence, which we call the
{\em Legendre dual divergence} of $D_F$. We have~:

\begin{lemma}
$D_F(\v{p}||\v{q})=F(\v{p})+F^*(\v{q}')-\dotproduct{\v{p}}{\v{q}'}=D_{F^*}(\v{q}'||\v{p}')$
\label{lem-dualdiv}
\end{lemma}

\begin{proof}
By Eq.~\ref{eq-div'},
$D_F(\v{p}||\v{q})=F(\v{p})-F(\v{q})-\dotproduct{\v{p}-\v{q}}{\v{q}'}$,
and, according to Eq.~\ref{eq-conjugate}, we have
$F(\v{p})=\dotproduct{\v{p}'}{\v{p}}-F^*(\v{p}')$ and
$F(\v{q})=\dotproduct{\v{q}'}{\v{q}}-F^*(\v{q}')$.  Hence,
$D_F(\v{p}||\v{q})=\dotproduct{\v{p}'}{\v{p}}-F^*(\v{p}')-\dotproduct{\v{p}}{\v{q}'}+F^*(\v{q}')=D_{F^*}(\v{q}'||\v{p}')$
since $\v{p}=\gradF^{-1}\gradF(\v{p})=\gradF^{*}(\v{p}')$.
\end{proof}

Observe that, when $D_F$ is symmetric, $D_{F^*}$ is also symmetric.

The Legendre transform of the quadratic form $F(\v{x})=\frac{1}{2}\v{x}^T\v{Q}\v{x}$,  where $\v{Q}$ is a symmetric invertible matrix, is $F^*(\v{y})=\frac{1}{2}\v{y}^T\v{Q}^{-1}\v{y}$ (corresponding divergences $D_F$ and $D_{F^*}$ are both generalized quadratic distances).

To compute $F^*$, we use the fact that $\gradFstar=\gradIF$ and
obtain $F^*$ as $F^*=\int \gradIF$. For example, the
Hellinger-like 
 measure is obtained by setting $F(x)=-\sqrt{1-x^2}$ (see
Table~\ref{tab:fbreg}).  The inverse gradient is
$\frac{x}{\sqrt{1+x^2}}$ and the dual convex conjugate is $\int
\frac{x\dx}{\sqrt{1+x^2}}=\sqrt{1+x^2}$. Integrating functions
symbolically may be difficult or even not possible, and, in some
cases, it will be required to approximate
numerically the inverse gradient $\gradIF(\v{x})$.

Let us consider the univariate generator functions defining the
divergences of Table~\ref{tab:fbreg}.  Both the  squared function
$F(\v{x})=x^2$ and Burg entropy $F(x)=-\log x$ are {\it self-dual},
i.e. $F=F^*$. This is easily seen by noticing that the gradient and
inverse gradient are identical (up to some constant factor).

For the exponential function $F(x)=\exp x$, we have $F^*(y) = y\log y-
y$ (the unnormalized Shannon entropy) and for the dual bit entropy $F(x)=\log(1+\exp x)$, we have $F^*(y) = y \log
\frac{y}{1-y} + \log (1-y)$, the bit entropy. 
Note that the bit entropy function is  a particular Bregman generator satisfying $F(x)=F(1-x)$.

\subsection{Symmetrized Bregman
divergences\label{sec:Bregman:SymmetrizedDivergence}}
For non-symmetric $d$-variate Bregman divergences $D_F$, we  define the {\it symmetrized divergence}

\[
S_F(\v{p},\v{q})=S_F(\v{q},\v{p})=\frac{1}{2} \left(
D_F(\v{p}||\v{q})+D_F(\v{q}||\v{p})\right) =\frac{1}{2} 
\dotproduct{\v{p}-\v{q}}{\v{p}'-\v{q}'} .
\]

An example of such a symmetrized divergence is the symmetric Kullback-Leibler divergence (SKL) widely used in computer vision and sound processing (see for example~\cite{entboost}).

A key observation is to note that the divergence $S_F$ between two points of $\set{X}$ can be measured as a  divergence in $\set{X}\times\set{X}'\subset\R^{2d}$.  More precisely, let ${\tilde\v{x}}=[\v{x}\ \v{x}' ]^T$ be the \hbox{$2d$-dimensional} vector obtained by stacking the coordinates of $\v{x}$ on top of those of $\xx '$, the gradient of $F$ at $\v{x}$. We have~:

\begin{theorem}
$S_F(\v{p},\v{q})=\frac{1}{2} D_{\tilde F}(\tilde \v{p} || \tilde \v{q})$  where $\tilde F(\tilde \v{x})= F(\v{x})+ F^*(\v{x}')$ and $D_{\tilde F}$ is the Bregman divergence defined over $\set{X}\times \set{X}'\subset \R^{2d}$ for the generator function $\tilde F$.
\end{theorem}

\begin{proof} Using Lemma~\ref{lem-dualdiv}, we have \[ S_F(\v{p},\v{q})=\frac{1}{2} \left(
D_F(\v{p}||\v{q})+D_F(\v{q}||\v{p})\right) = \frac{1}{2} \left( D_F(\v{p}||\v{q})+D_{F^*}(\v{p}'||\v{q}')\right)
= \frac{1}{2} D_{\tilde F}(\tilde \v{p} || \tilde \v{q}) \] \end{proof}

It should be noted that $\tilde\xx $ lies on the 
$d$-manifold  $\tilde\set{X}=\{ [\v{x}\
\v{x}']^T \ |\
\v{x}\in\mathbb{R}^d\}$ of $\mathbb{R}^{2d}$.  Note also that $S_F(\v{p},\v{q})$  is symmetric but {\em not} a Bregman divergence in general since $\tilde\set{X}$ may not be convex, while $D_{\tilde F}$ is a non symmetric Bregman divergence in $\set{X}\times\set{X}'$.

\subsection{Exponential families\label{sec:Bregman:ExponentialFamilies}}

\subsubsection{Parametric statistical spaces and exponential families}

A {\em statistical space} $\set{X}$ is an abstract space where coordinates of vector points $\thetav\in\set{X}$ encode the parameters of  statistical distributions. The dimension $d=\dim \set{X}$ of the statistical space coincides with the finite number of free parameters of the distribution laws.
For example, the space $\set{X}=\{[\mu\ \sigma]^T\ |\ (\mu,\sigma)\in\mathbb{R}\times\mathbb{R}^+_*\}$ of univariate normal distributions $\set{N}(\mu,\sigma)$ is a 2D parametric statistical space, extensively studied in information geometry~\cite{informationgeometry} under the auspices of differential geometry.
A prominent class of distribution families called the {\em exponential families} $\set{E}_F$~\cite{informationgeometry} admits the same {\em canonical} probability distribution function

\begin{equation}\label{eq:expofamily}
p(x|\thetav)\equaldef\exp\{\dotproduct{\thetav}{\v{f}(x)} - F(\thetav) + C(x)\}, \end{equation}

\noindent where $\v{f}(x)$ denotes the {\em sufficient statistics} and $\thetav\in\set{X}$ represents the {\it natural parameters}.
Space $\set{X}$ is thus called the {\it natural parameter space} and, since $\log \int_{x}  p(x|\thetav) \dx=\log 1=0$, we have $F(\thetav)=\log\int_{x} \exp\{\dotproduct{\thetav}{\v{f}(\v{x})} +
C(x) \} \dx$. $F$ is called the {\it cumulant function} or the {\em log-partition} function.  $F$ fully characterizes the exponential family $\set{E}_F$ while term $C(x)$ ensures density normalization.
(That is, $p(x|\thetav)$ is indeed a probability density function satisfying $\int_{x} p(x| \thetav) \dx=1$.)

\begin{table}
\renewcommand{\arraystretch}{1.5}
$$
\begin{array}{|l|l|l|l|} \hline
\multicolumn{4}{|c|}{\mbox{\large Exponential family}} \\ \multicolumn{4}{|c|}{\mbox{Canonical probability density function:} \exp\{\dotproduct{\thetav}{\v{f}(x)} - F(\thetav) + C(x)\}} \\ \hline \hline

\hline
 \mbox{Natural} & \mbox{Sufficient}  & \mbox{Cumulant function}\ F(\thetav)  & \mbox{Dens. Norm.}\ \\
 \mbox{parameters}\ \thetav  & \mbox{statistics}\ \vector{f}(x)&   & 
C(x)\\
\hline\hline
\multicolumn{4}{|c|}{\mbox{Bernouilli $\set{B}(q)$ (Tossing coin with $\Pr(\mathrm{heads})=q$ and $\Pr(\mathrm{tails})=1-q$) }} \\ \hline
 \log\frac{q}{1-q} & x & \log(1+\exp\theta) & 0 \\ \hline\hline
  \multicolumn{4}{|c|}{ \mbox{ Multinomial $\set{M}(q_1, ..., q_{d+1})$ (Extend Bernouilli with $\Pr(x_i)=q_i$ and $\sum_i q_i=1$)}}\\ \hline 
   \theta_i=\log \frac{q_i}{1-\sum_{j=1}^{d}q_i} & f_i(\v{x})=x_i & \log (1+\sum_{i=1}^d \exp\theta_i) & 0\\ \hline
 \multicolumn{4}{|c|}{\mbox{Beta $\beta(\theta_1, \theta_2)$ (Bernouilli conjugate prior) }}\\ \hline
 [\theta_1\ \theta_2]^T & [\log x\ \log(1-x)]^T & \log B(\theta_1+1,\theta_2+1) & 0 \\ \hline
 \multicolumn{4}{|c|}{ F(\thetav)= \log\frac{\Gamma(\theta_1+1)\Gamma(\theta_2+1)}{\Gamma(\theta_1+\theta_2-2)} }\\
  \multicolumn{4}{|c|}{ (\mbox{with}\ \Gamma(x)=\int_0^\infty t^{x-1}\exp(-t)\mathrm{d}t)=(x-1)\Gamma(x-1) )}\\ \hline\hline
  \multicolumn{4}{|c|}{ \mbox{Univariate Normal $\set{N}(\mu,\sigma^2)$} }\\ \hline
 [\frac{\mu}{\sigma^2}\ \frac{-1}{2\sigma^2} ]^T & [x\ x^2]^T & -\frac{\theta_1^2}{4\theta_2}+\frac{1}{2}\log (-\frac{\pi}{\theta_2})  & 0\\ \hline\hline
 \multicolumn{4}{|c|}{\mbox{Multivariate Normal $\set{N}(\muv,\Sigmam)$ }}\\ \hline
[\Sigmam^{-1}\muv\ -\frac{1}{2}\Sigmam^{-1}] & [\v{x}\ \v{x}\v{x}^T ]  & \frac{1}{2}\muv^T\Sigmam^{-1}\muv+\frac{1}{2}\log \det (2\pi\Sigmam) & 0 \\ \hline\hline 
 \multicolumn{4}{|c|}{\mbox{Rayleigh $\set{R}(\sigma^2)$ (used in ultrasound imageries)}}\\ \hline 
-\frac{1}{2\sigma^2} & x^2 & \log -\frac{1}{2\theta}  & \log x \\ \hline\hline
 \multicolumn{4}{|c|}{\mbox{Laplacian $\set{L}(\theta)$ (used in radioactivity decay) }}\\ \hline
 \theta & -x & -\log \theta & 0\\ \hline\hline
  \multicolumn{4}{|c|}{\mbox{Poisson $\set{P}(\lambda)$ (counting process) }}\\ \hline
  \log\lambda & x & \exp\theta & -\log x!\\ \hline \hline
  \multicolumn{4}{|c|}{\mbox{Gamma $\gamma(\theta_1,\theta_2)$ (waiting times in Poisson processes)}}\\ \hline
 [\theta_1\ \theta_2]^T & [\log x\ x]^T & \log\Gamma(\theta_1+1)+(\theta_2+1)\log (-\theta_2) & 0 \\ \hline \hline
 \multicolumn{4}{|c|}{\mbox{Dirichlet $\set{D}(\alphav)$ (varying proportion model $||\v{x}||=1$, conjugate prior of Multinomial)}} \\ \hline
 \theta_i=\alpha_i-1 & f_i(\v{x})=\log x_i & \log\Gamma(\sum_i\theta_i+d)-\sum_i\Gamma(\theta_i+1) & 0 \\ \hline
  \end{array} $$ 
  \caption{Canonical decompositions of usual exponential families.\label{tab:expo}} \end{table}

When the components of the sufficient statistics are affinely
independent, this canonical representation is said to be {\em minimal}, and
the family $\set{E}_F$ is called a {\it full} exponential family of
order $d=\dim\set{X}$.  Moreover, we consider {\em regular}
exponential families $\set{E}_F$ that have their support domains
topologically open.  Regular exponential families include many famous
distribution laws such as Bernoulli (multinomial), Normal (univariate, multivariate and rectified), Poisson, Laplacian, negative binomial, Rayleigh, Wishart, Dirichlet, and Gamma distributions.  Table~\ref{tab:expo}
summarizes the various relevant parts of the canonical decompositions
of some of these usual statistical distributions. 
Observe that the product of any two distributions of the same exponential family is another exponential family distribution that may not have anymore a nice parametric form (except for products of normal distribution pdfs that yield again normal distribution pdfs).
Thus exponential families provide a unified treatment framework of common distributions.
 Note, however, that the uniform
distribution {\em does not} belong to the exponential families.

\subsubsection{Kullback-Leibler divergence of exponential families\label{sec:kl}}

In such statistical spaces $\set{X}$, a basic primitive is to measure the {\em distortion} between any two distributions.
The {\em Kullback-Leibler divergence} (also called {\em relative
entropy} or information divergence, $I$-divergence)  is a standard information-theoretic measure between two  
statistical distributions $d_1$ and $d_2$   defined as
$
\KL(d_1||d_2)\equaldef\int_x d_1(x)\log \frac{d_1(x)}{d_2(x)}\dx$.
This statistical measure is not symmetric nor does the triangle inequality holds.

The link with Bregman divergences comes from the remarkable property that the Kullback-Leibler divergence of any two distributions of the {\it same} exponential family with respective natural parameters $\thetav_p$ and $\thetav_q$ is obtained from the Bregman divergence induced by the cumulant function of that family by {\it swapping} arguments.
By a slight abuse of notations, we denote by $\KL(\thetav_p||\thetav_q)$  the oriented Kullback-Leibler divergence between the probability density functions defined by the respective natural parameters, i.e. $\KL(\thetav_p||\thetav_q)\equaldef\int_x p(x|\thetav_p)\log \frac{p(x|\thetav_p)}{p(x|\thetav_q)}\dx$. 
The following theorem is the extension to the continuous case of a result
mentioned in ~\cite{j-cbd-2005}.

\begin{theorem}\label{KLBD}
The Kullback-Leibler divergence of any two distributions of the  same exponential family with  natural parameters $\thetav_p$ and $\thetav_q$ is obtained from the Bregman divergence induced by the cumulant function $F$ as:  $\KL(\thetav_p||\thetav_q)=D_F(\thetav_q||\thetav_p)$.
\end{theorem}

Before proving the theorem, we note that
\begin{equation}
\grad F(\thetav)=\left[\int_x \v{f}(x) \exp\{\dotproduct{\thetav}{\v{f}(x)} - F(\thetav) +C(x)\} \dx\right].\label{eq-expo-grad}
\end{equation}

The coordinates of
$\muv\equaldef\grad F(\thetav)=[\int _x
\v{f}(\v{x})p(x|\thetav)\dx]=E_{\thetav}(\v{f}(\v{x}))$ are called the
{\it expectation parameters}. As an example, consider the univariate normal
distribution $\set{N}(\mu,\sigma)$ with sufficient statistics $[x\
x^2]^T$ (see Table~\ref{tab:expo}). The expectation parameters are
$\muv=\grad_F(\thetav)=[\mu\ \mu^2+\sigma^2]^T$, where $\mu=\int _x x\,
p(x|\thetav)\dx$ and $\mu^2+\sigma^2=\int _x x^2 p(x|\thetav)\dx$.

We now prove the theorem.

\begin{proof}
\begin{eqnarray*}
\KL(\thetav_p||\thetav_q)
&=& \int_x p(x|\thetav_p)\log
\frac{p(x|\thetav_p)}{p(x|\thetav_q)} \dx\\ &=& \int_x p(x|\thetav_p)
(F(\thetav_q)-F(\thetav_p)+\dotproduct{\thetav_p-\thetav_q}{\v{f}(x)})
\dx \\
&=& \int_x p(x|\thetav_p)  \left
(D_F(\thetav_q||\thetav_p)+\dotproduct{\thetav_{{q}}-\thetav_{{p}}}{\gradF(\thetav_p)}+\dotproduct{\thetav_p-\thetav_q}{\v{f}(x)}\right)
\dx \\
&=& D_F(\thetav_q||\thetav_p)+ \int_x p(x|\thetav_p)
\dotproduct{\thetav_q-\thetav_p}{\gradF(\thetav_p)-\v{f}(x)}) \dx \\ &=& D_F(\thetav_q||\thetav_p)  - \int_x p(x|\thetav_p) \dotproduct{\thetav_q-\thetav_p}{\v{f}(x)} \dx + \dotproduct{\thetav_q-\thetav_p}{\gradF(\thetav_p)}\\
&\stackrel{({\rm Eq.}~\ref{eq-expo-grad})}{=}& D_F(\thetav_q||\thetav_p)
\end{eqnarray*}
\end{proof}

\subsubsection{Dual parameterizations and dual divergences} 

The notion of dual Bregman divergences introduced earlier and dual
parameterizations extend naturally to statistical spaces.  Since,
$\muv= \grad F(\thetav)$ (Eq.~\ref{eq-expo-grad}), the
convex conjugate of $F(\thetav)$ is
$F^*(\muv)=\dotproduct{\thetav}{\muv}-F(\thetav)$
(Eq.~\ref{eq-conjugate}). From Lemma~\ref{lem-dualdiv}, we then deduce
the following theorem.

\begin{theorem}
$D_F(\thetav_p||\thetav_q)=D_{F^*}(\muv_q||\muv_p)$ 
where $F^*$ denote the convex conjugate of $F$.
\end{theorem}

Table~\ref{tab:expfamilyduality} presents some examples of dual parameterizations of exponential families (i.e., the natural $\thetav$-parameters and expectation $\muv$-parameters and dual Legendre cumulant functions), and describe the corresponding Bregman divergences induced by the Kullback-Leibler divergences.

\begin{table}
\centering
\renewcommand{\arraystretch}{1.5}
\footnotesize
$$
\begin{array}{|l|l|l|} \hline
\multicolumn{3}{|c|}{\mbox{Bernouilli dual divergences: Logistic loss/binary relative entropy}}\\ \hline
F(\theta)=\log(1+\exp\theta) & D_F(\theta||\theta')=\log \frac{1+\exp\theta}{1+\exp\theta'}-(\theta-\theta')\frac{\exp\theta'}{1+\exp\theta'}
& f(\theta)=\frac{\exp\theta}{1+\exp\theta}=\mu \\ \hline F^*(\mu)=\mu \log\mu+(1-\mu)\log (1-\mu) & D_{F^*}(\mu'||\mu)=\mu'\log\frac{\mu'}{\mu}+(1-\mu)\log\frac{1-\mu'}{1-\mu}
& f^*(\mu)=\log\frac{\mu}{1-\mu}=\theta \\ \hline \hline \multicolumn{3}{|c|}{\mbox{Poisson dual divergences: Exponential loss/Unnormalized Shannon entropy}}\\ \hline F(\theta)=\exp\theta & D_F(\theta || \theta')=\exp\theta-\exp\theta'-(\theta-\theta')\exp\theta' & f(\theta)=\exp\theta=\mu\\ \hline F^*(\mu)=\mu\log\mu-\mu & D_{F^*}(\mu'||\mu)= \mu'\log\frac{\mu'}{\mu}+\mu-\mu' & f^*(\mu)=\log \mu=\theta \\ \hline \end{array} $$ \caption{Examples of dual parameterizations of exponential families and their corresponding Kullback-Leibler (Bregman) divergences for the Bernoulli and Poisson distributions.\label{tab:expfamilyduality}}
\end{table}

Finally, we would like to point out that Banerjee et al.~\cite{j-cbd-2005} have shown that there is a {\it bijection} between the regular exponential families and a subset of the Bregman divergences called {\em regular Bregman divergences}.

\section{Elements of Bregman geometry}
\label{sec-bregman-geometry}

In this section, we discuss several basic geometric properties that
will be useful when studying Bregman Voronoi diagrams.  Specifically, we
characterize Bregman bisectors, Bregman balls and Bregman
geodesics. Since Bregman divergences are not symmetric, we describe
several types of Bregman bisectors in
\S\ref{sec:BVD:bisector}. We subsequently  characterize Bregman balls
by using a lifting transform that extends a construction well-known in the
Euclidean case (\S\ref{sec:SpherePolarity}).  Finally, we  characterize
geodesics and show an orthogonality property between bisectors and
geodesics in
\S\ref{sec:BVD:geodesicbisector}.

\subsection{Bregman bisectors\label{sec:BVD:bisector}}

Since Bregman divergences are not symmetric, we can define several
types of bisectors. The Bregman bisector of the {\em first type} is
defined as 
$$H_F({\v{p}},{\v{q}})=\{
\v{x}\in\set{X}\ |\
D_F(\v{x}||\v{p})=D_F(\v{x}||\v{q})\}.$$  
Similarly, we define the
Bregman bisector of the {\em second type } as 
$$H'_F({\v{p}},{\v{q}})=\{
\v{x}\in\set{X}\ |\
D_F(\v{p}||\v{x})=D_F(\v{q}||\v{x})\}.$$
These bisectors are identical
when the divergence is symmetric. However, in general, they are
distinct, the bisectors of the first type being linear while the
bisectors of the second type are potentially curved (but always linear in the gradient space, hence the notation). More precisely, we have the
following  lemma

\begin{lemma}\label{fsbb}
The  Bregman bisector of the first type $H_F(\v{p},\v{q})$ is the hyperplane of equation: $$
H_F({\v{p}},{\v{q}}): \dotproduct{\vector{x}}{\v{p}'-\v{q}'} + F(\v{p}) - \dotproduct{\v{p}}{\v{p}'} - F(\v{q}) + \dotproduct{\v{q}}{\v{q}'} =  0\ 
$$ 
The Bregman bisector of the second type $H'_F(\v{p},\v{q})$ is 
the hypersurface of equation
$$H'_F(\v{p},\v{q}):\dotproduct{\v{x}'}{\v{q}-\v{p}}+F(\v{p})-F(\v{q})=0
$$ 
(a hyperplane in the gradient space $\set{X}'$).
\end{lemma}

It should be noted that $\v{p}$ and  $\v{q}$ lie necessarily on different sides of
$H_F(\v{p},\v{q})$ since 
$H_F(\v{p},\v{q})(\pp)=-D_F(\pp || \qq)<0$ and $H_F(\v{p},\v{q})(\qq)= D_F(\qq ||\pp)>0$.

From Lemma \ref{lem-dualdiv}, we know that 
$D_F(\v{x}||\v{y})= D_{F^*}(\v{y}'||\v{x}')$
where $F^*$ is the convex conjugate of $F$.
We therefore have
\begin{eqnarray*}
H_F({\v{p}},{\v{q}}) &=& \gradIF (H'_{F^*}({\v{q}'},{\v{p}'})),\\
H'_{F}({\v{p}},{\v{q}}) &=&  \gradIF (H_{F^*}({\v{q}'},{\v{p}'})).
\end{eqnarray*}

\begin{figure}
\centering
\def\ttt{5.5cm}
\begin{tabular}{ccc}
& Source space $\set{X}$ & Gradient space $\set{X}'$\\
(a) & \includegraphics[bb=0 0 799 799 , width=\ttt]{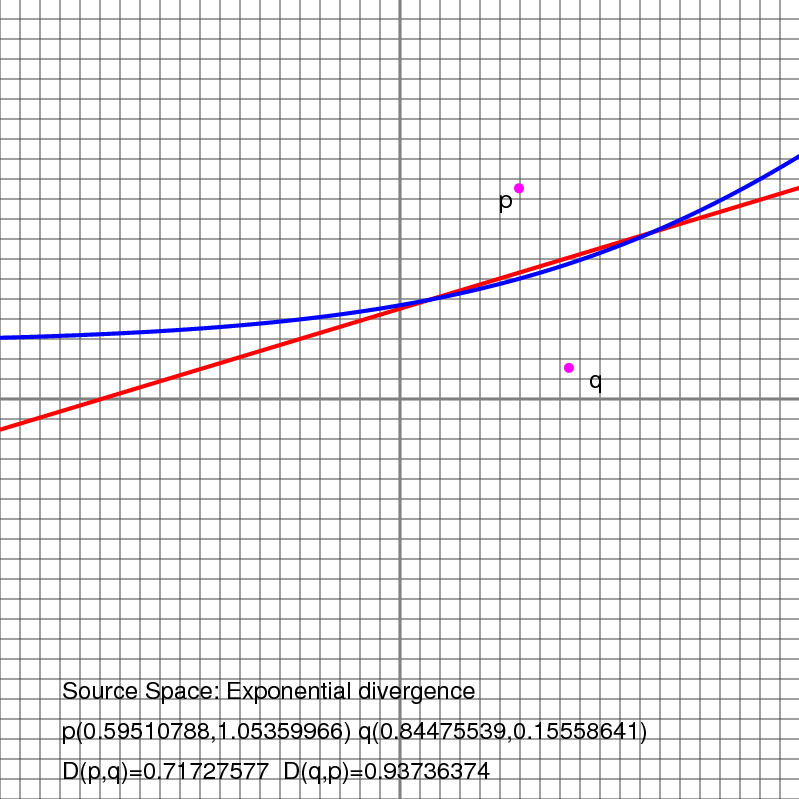} & \includegraphics[bb=0 0 799 799 , width=\ttt]{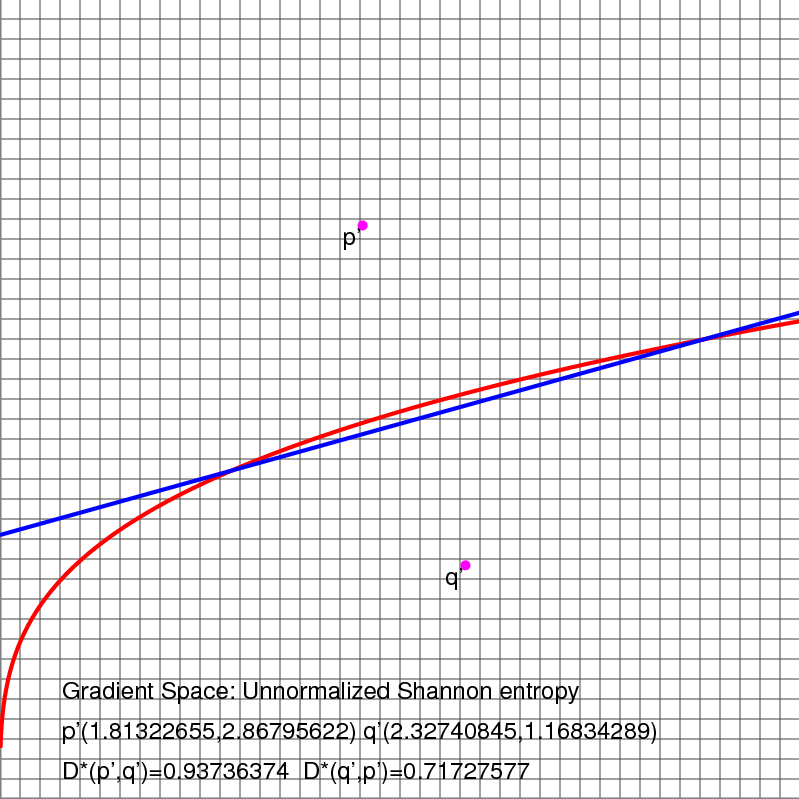} \\
(b) &\includegraphics[bb=0 0 799 799 , width=\ttt]{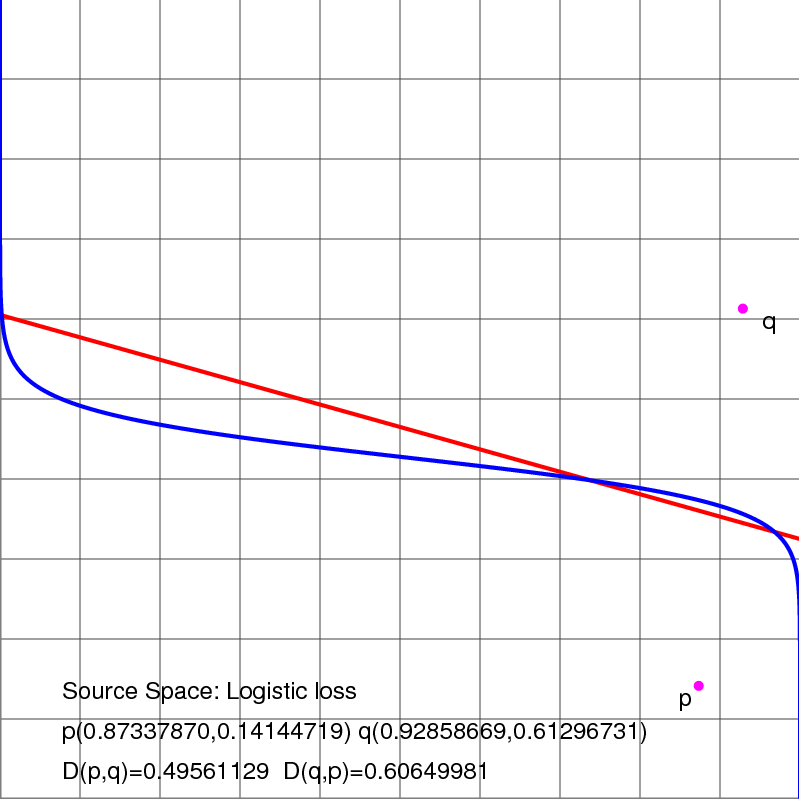} & \includegraphics[bb=0 0 799 799 , width=\ttt]{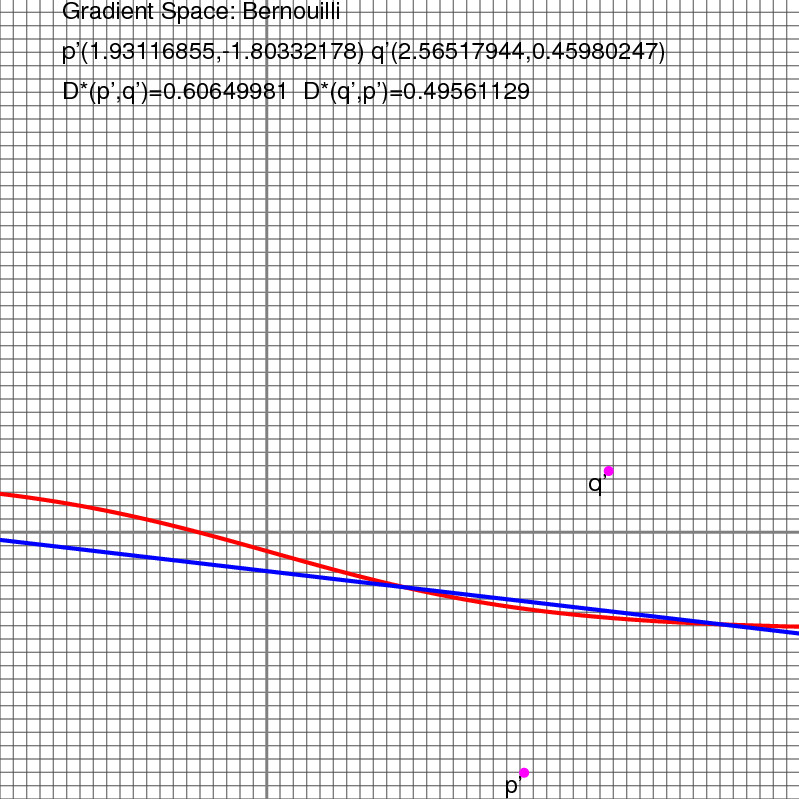} \\
(c) & \includegraphics[bb =0 0 799 799 , width=\ttt]{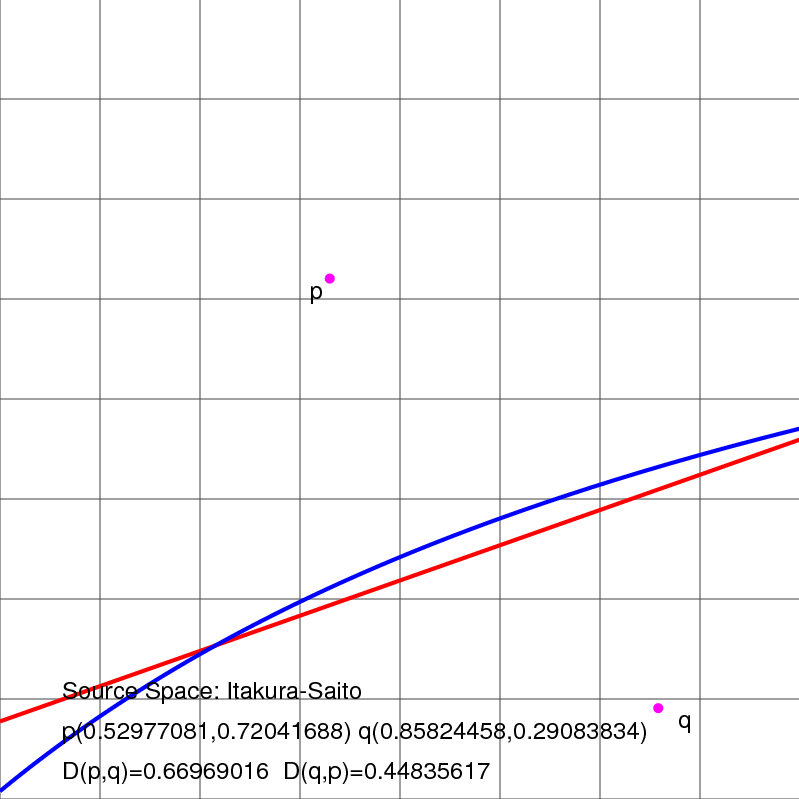} & \includegraphics[bb=0 0 799 799 , width=\ttt]{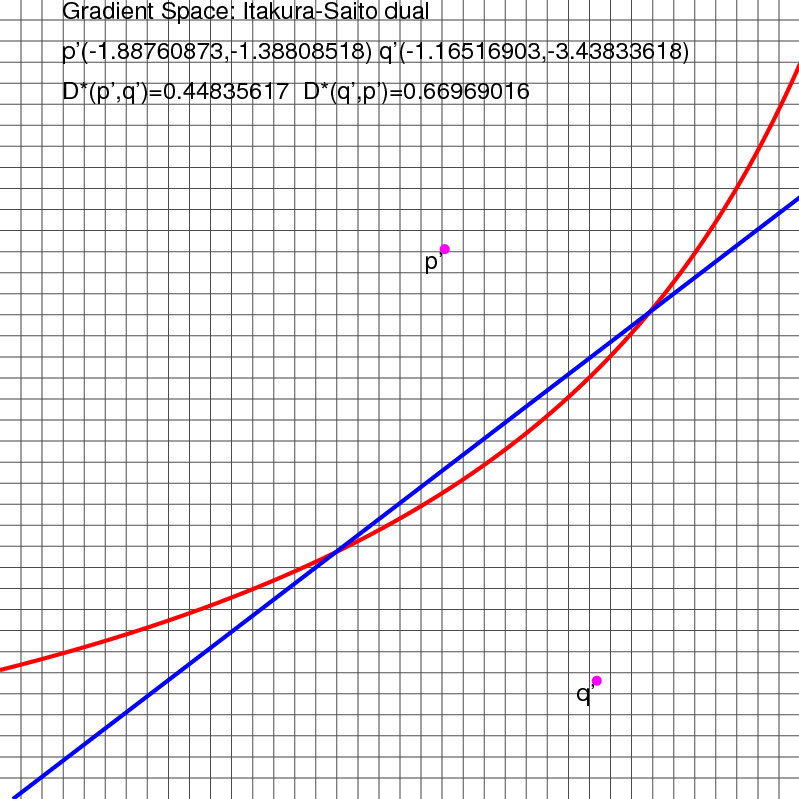}
\end{tabular}

\caption{Bregman bisectors: first-type linear bisector  and second-type curved bisector  are displayed for pairs of primal/dual Bregman divergences:  (a) exponential loss/unnormalized Shannon entropy,  (b)  logistic loss/dual logistic loss, and  (c)  self-dual Itakura-Saito divergence. The grid size of $\mathbb{R}^2$ in $\set{X}$ and $\set{X}'$ is ten ticks per unit. First-type (primal linear/dual curved) and second-type (primal curved/dual linear) bisectors are respectively drawn in red and blue. \label{fig:ttbisectors}}
\end{figure}

Figure~\ref{fig:ttbisectors} depicts several first-type and
second-type bisectors for various pairs of primal/dual Bregman divergences.

The bisector $H^{''}_F(\v{p},\v{q})$ for the {\em symmetrized
Bregman divergence} $S_F$ is given by
\[
H^{''}_F({\v{p}},{\v{q}}):
\dotproduct{\v{x}}{\v{q}'-\v{p}'}+\dotproduct{\v{x}'}{\v{q}-\v{p}}+\dotproduct{\v{p}}{\v{p}'}-\dotproduct{\v{q}}{\v{q}'}=0.
\]
Such a bisector is not linear in $\v{x}$ nor  in $\v{x}'$.
However, we can observe that the expression is linear in
$\tilde\v{x}=[\v{x}\ \v{x}' ]^T$.  Indeed, proceeding as we did in
\S\ref{sec:Bregman:SymmetrizedDivergence}, we can rewrite the above 
equation as
\[
H_{\tilde{F}}(\tilde{\pp},\tilde{\qq}):\Dotproduct{\twovector{\v{x}}{\v{x}'} }{
\twovector{\v{q}'-\v{p}'}{\v{q}-\v{p}}
}+\dotproduct{\v{p}}{\v{p}'}-\dotproduct{\v{q}}{\v{q}'}=0.
\]
which shows that $H^{''}_F(\v{p},\v{q})$ is the projection on $\set{X}$ of the
intersection of the hyperplane $H(\tilde{\pp},\tilde{\qq})$ of $\R^{2d}$
with the  $d$-dimensional manifold
$\tilde\set{X}= \{ {\tilde\v{x}}=[\v{x}\ \v{x}']^T\ |\  \v{x}\in \set{X}\}$.

\subsection{Bregman spheres and the lifting map\label{sec:SpherePolarity}}

We  define the Bregman balls of, respectively,  the first and the second 
types as
\[
\ball(\v{c},r)=\{ \v{x}\in\set{X}\ |\ D_F(\v{x}||\v{c})\leq r\}
\quad {{\rm and}} \quad \ball'(\v{c},r)=\{ \v{x}\in\set{X}\ |\ D_F(\v{c}||\v{x})\leq r\}
\]

\begin{figure}
\centering
\begin{tabular}{ccc}
\includegraphics[bb=0 0 399 399 , width=5cm]{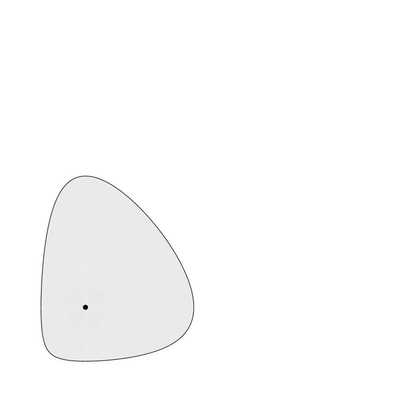} &
\includegraphics[bb=0 0 399 399  , width=5cm]{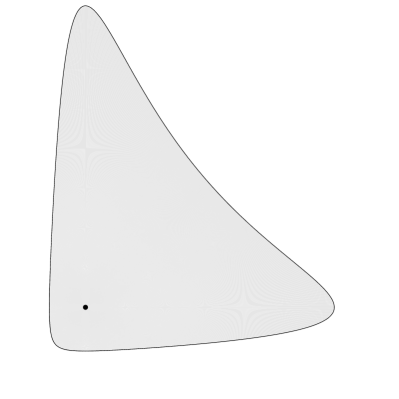} &
\includegraphics[bb=0 0 399 399  , width=5cm]{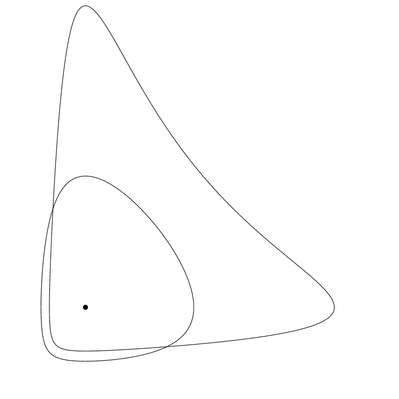}  \\
(a) & (b) & (c)
\end{tabular}
\caption{Bregman balls for the Itakura-Saito divergence. The (convex) 
ball~(a) of the first type $\ball(\v{c},r)$, (b) the ball of the second
type $\ball'(\v{c},r)$ with the same center and radius, (c)
superposition of the two corresponding bounding spheres.
\label{fig:convexballs} }
\end{figure}

The Bregman balls of the first type are convex while this is not
necessarily true for the balls of the second type as shown in
Fig.~\ref{fig:convexballs} for the Itakura-Saito divergence (defined in Table~\ref{tab:fbreg}). The
associated bounding {\em Bregman spheres} are obtained by replacing
the inequalities by equalities.

From Lemma~\ref{lem-dualdiv}, we deduce that 
\begin{equation}
\ball'(\v{c},r)=\gradIF (\balll(\v{c}',r)).
\label{eq-bballs}
\end{equation}

Let us now examine a few properties of Bregman spheres using a lifting
transformation that generalizes a similar
construct for Euclidean spheres
(see~\cite{compgeom-1998,inria-recherche-1620}).

Let us embed the domain $\set{X}$ in
$\hat\set{X}=\set{X}\times\mathbb{R}\subset \R^{d+1}$ using an {\it extra dimension}
denoted by the $Z$-axis.   For a point
$\vector{x}\in\mathcal{X}$, recall that 
$\hat{\vector{x}}=(\vector{x},F(\vector{x}))$ denotes the point
obtained by lifting $\v{x}$ onto $\mathcal{F}$ (see Figure~\ref{fig:ordinaryl22}).
In addition, write $\proj_{\set{X}}(\vector{x},z)=\vector{x}$ for the
projection of a point of $\hat\set{X}$ onto $\set{X}$.

Let $\v{p}\in \set{X}$ and $H_{\v{p}}$ be the  hyperplane tangent to 
$\mathcal{F}$
at point $\hat{\vector{p}}$ of equation
$$
z= H_{\v{p}}(\vector{x})=\dotproduct{\vector{x}-\vector{p}}{\vector{p}'}+F(\vector{p}), 
$$
and let $H_{\v{p}}^{\uparrow}$ denote the halfspace  above $H_{\v{p}}$
consisting of the  points
$\v{x}=[\v{x}\ z]^T\in\hat\set{X}$ such that
$z>H_{\v{p}}(\vector{x})$. 
Let $\sigma(\v{c},r)$ denote either the first-type or second-type Bregman sphere centered at $\v{c}$ with radius $r$ (i.e., $\partial B_F(\v{c},r)$ or $\partial B_F'(\v{c},r)$).

\begin{figure}
\centering
\begin{tabular}{cc}
\includegraphics[bb=0 0 511 511,width=7cm]{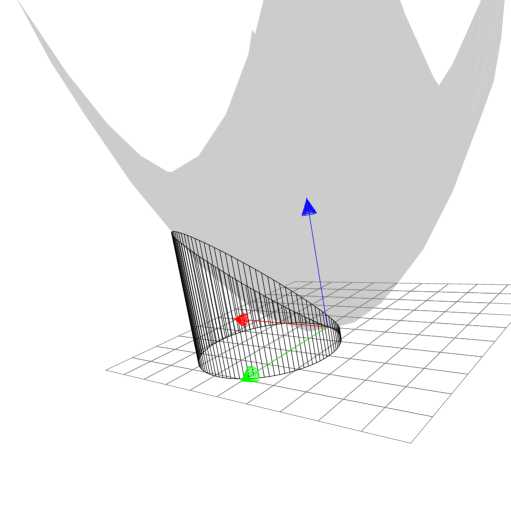} &
\includegraphics[bb=0 0 511 511,width=7cm]{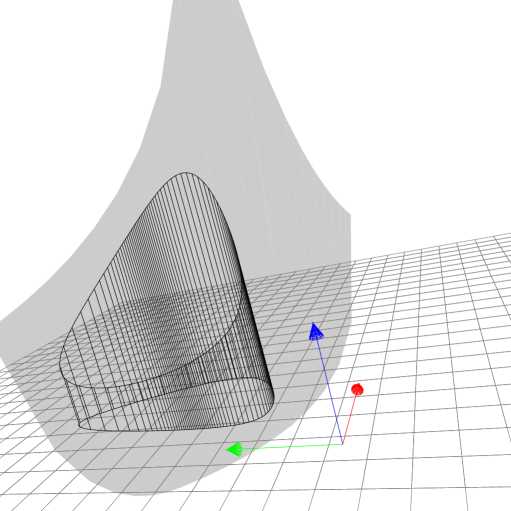} \\
(a) Squared Euclidean distance & (b) Itakura-Saito divergence
\end{tabular}
\caption{Two Bregman circles $\sigma$ and the associated curves $\hat{\sigma}$ obtained by lifting $\sigma$ onto $\mathcal{F}$. 
The curves $\hat{\sigma}$ are obtained as the intersection of the hyperplane $H_\sigma$ with the  convex hypersurface $\set{F}$.
3D illustration with (a) the squared Euclidean distance, and (b) the Itakura-Saito divergence.
\label{fig:sphereplane}}
\end{figure}

The lifted image $\hat{\sigma}$ of a Bregman sphere $\sigma$ is
$\hat{\sigma}= \{ (\v{x},F(\v{x})), \v{x}\in \sigma\}$.
We associate to a Bregman sphere $\sigma = \sigma (\v{c},r)$ of
$\mathcal{X}$ the hyperplane
\begin{equation}
H_{\sigma} : z= \dotproduct{\v{x}-\v{c}}{\v{c}'}+ F(\v{c}) +r,
\label{eq-Hsigma}
\end{equation}
parallel to $H_{\v{c}}$ and at vertical distance $r$ from $H_{\v{c}}$ (see Figure~\ref{fig:sphereplane}).
Observe that $H_{\sigma}$ coincides with $H_{\v{c}}$ when $r=0$,
i.e. when sphere $\sigma$ is reduced to a single point.

\begin{lemma}
\label{lem-lift-sphere}
$\hat{\sigma}$ is the intersection of $\mathcal{F}$ with $H_{\sigma}$.
Conversely, the intersection of any hyperplane $H$ with $\set{F}$
projects onto $\set{X}$ as a Bregman sphere. More precisely, if
the equation of $H$ is $z= \dotproduct{\v{x}}{\v{a}} + b$, the
sphere is centered at $\v{c}=\gradIF (\v{a})$ and its radius is 
$\dotproduct{\v{a}}{\v{c}} -F(\v{c}) + b$.
\end{lemma}

\begin{proof}
The first part of the lemma is a direct consequence of 
the fact that $D_F(\v{x}||\v{y})$ is measured by the vertical distance
from $\hat{\v{x}}$ to $H_{\v{y}}$ (see Lemma~\ref{lem-code-BD}).
For the second part, we consider the hyperplane $H^{\|}$ 
parallel to $H$ and tangent to $\mathcal{F}$. From Eq.~\ref{eq-Hsigma}, we deduce
$\v{a}=\v{c}'$.  The equation of $H^{\|}$ is thus $z =
\dotproduct{\v{x}-\gradIF (\v{a})}{\v{a}}+ F(\gradIF
(\v{a}))$.  It follows that the divergence from any point of $\sigma$
to $\v{c}$, which is equal to the vertical distance between $H$ and
$H^{\|}$, is $\dotproduct{\gradIF (\v{a})}{\v{a}}- F(\gradIF (\v{a}))
+b= \dotproduct{\v{a}}{\v{c}} -F(\v{c}) + b$.
\end{proof}

Bregman spheres have been defined as manifolds of codimension 1 of $\R
^d$, i.e. hyperspheres. More generally, we can define the Bregman
spheres of codimension $k+1$ of $\R ^d$ as the Bregman (hyper)spheres
of some affine space $\mathcal{Z}\subset \R^d$ of codimension $k$. The
next lemma shows that Bregman spheres are stable under intersection.

\begin{lemma}\label{intersec}
The intersection of $k$ Bregman spheres $\sigma_1,\ldots ,\sigma_k$ is
a Bregman sphere $\sigma$. If the $\sigma_i$ pairwise intersect
transversally, $\sigma=\cap_{i=1}^k \sigma_i$ is a $k$-Bregman sphere. 
\end{lemma}

\begin{proof}
Consider first the case of Bregman spheres of
the first type.  The $k$ hyperplanes $H_{\sigma_i}$, $i=1,\ldots ,k$
intersect along an affine space $H$ of codimension $k$ of $\R ^{d+1}$
that vertically projects onto $G$. Let $G^{\updownarrow}=G\times \R$
be the vertical flat of codimension $k$ that contains $G$ (and $H$)
and write $\mathcal{F}_G=\mathcal{F}\cap G^{\updownarrow}$ and $H_G=
H\cap G^{\updownarrow}$. Note that $\mathcal{F}_G$ is the graph of the
restriction of $F$ to $G$ and that $H_G$ is a hyperplane of
$G^{\updownarrow}$.  We can therefore apply
Lemma~\ref{lem-lift-sphere} in $G^{\updownarrow}$, which proves the
lemma for Bregman spheres of the first type.

The case of Bregman spheres of the second type  follows from the duality of Eq.~\ref{eq-bballs}.
\end{proof}

\paragraph{Union and intersection of Bregman balls}

\begin{theorem}
The union of $n$ Bregman balls has combinatorial complexity $\Theta
(n^{\floor{\frac{d+1}{2}}})$ and can be computed in time $\Theta
(n\log n+ n^{\floor{\frac{d+1}{2}}})$.
\end{theorem}

\begin{proof}
To each ball, we can associate its bounding Bregman sphere $\sigma_i$
which, by Lemma \ref{lem-lift-sphere}, is the projection by
$\proj_{\set{X}}$ of the intersection of $\mathcal{F}$ with a
hyperplane $H_{\sigma_i}$. The points of $\mathcal{F}$ that are below
$H_{\sigma_i}$ projects onto points that are inside the Bregman ball
bounded by $\sigma_i$. Hence, the union of the balls is the projection
by $\proj_{\set{X}}$ of the complement of $\mathcal{F}\cap
\mathcal{H}^{\uparrow}$ where $\mathcal{H}^{\uparrow}= \cap_{i=1}^n
H_{\sigma_i}^{\uparrow}$.  $\mathcal{H}^{\uparrow}$ is a convex
polytope defined as the intersection of $n$ half-spaces.  The theorem
follows from McMullen's theorem that bounds the
number of faces of a polytope~\cite{McMullen1971}, and Chazelle's optimal
 convex hull/half-space intersection algorithm~\cite{Chazelle1993}. The result for the balls of the second
type is deduced from the result for the balls of the first type and the duality of 
Eq.~\ref{eq-bballs}.
\end{proof}

Very similar arguments prove the following theorem (just replace
$H_{\sigma_i}^{\uparrow}$ by the complementary halfspace $H_{\sigma_i}^{\downarrow}$).

\begin{theorem}
The intersection of $n$ Bregman balls has combinatorial complexity
$\Theta (n^{\floor{\frac{d+1}{2}}})$ and can be computed in time
$\Theta (n\log n+ n^{\floor{\frac{d+1}{2}}})$.
\end{theorem}

\paragraph{Circumscribing Bregman spheres.}
There exists, in general, a unique Bregman sphere passing through
$d+1$ points of $\R ^d$. This is easily shown using the lifting map
since, in general, there exists a unique hyperplanes of $\R^{d+1}$ 
passing through $d+1$ points. The claim then follows from Lemma~\ref{lem-lift-sphere}.

Deciding whether a point $\v{x}$ falls {\em inside}, {\em on} or {\em outside}
a Bregman sphere $\sigma\in\mathbb{R}^d$ passing through $d+1$ points
of $\v{p}_0, ...,\v{p}_d$ will be crucial for computing Bregman
Voronoi diagrams and associated triangulations.  The lifting map immediately
implies that such a decision task reduces to determining the orientation of
the simplex $(\v{\hat{p}}_0, ...,\v{\hat{p}}_d,\hat{\xx})$ of $\R^{d+1}$,
which in turn reduces to evaluating the sign of the determinant of the
$(d+2)\times (d+2)$ matrix (see~\cite{b-n-visualcomputing-2005})

\[
\texttt{InSphere}(\v{x};\v{p}_0, ..., \v{p}_d) = \left| \begin{array}{cccc} 1 & ...
&1 & 1\\ \v{p}_0 & ... & \v{p}_{d} & \xx \\ F(\v{p}_0) & ... & F(\v{p}_{d})& F(\xx ) \end{array} \right|
\]
If one assumes that the determinant $\left| \begin{array}{ccc} 1 & ...
& 1\\ \v{p}_0 & ... & \v{p}_{d} \end{array}\right|$ is non-zero,
$\texttt{InSphere}(\v{x};\v{p}_0, ..., \v{p}_d)$ is negative, null or
positive depending on whether $\xx$ lies inside, on, or outside $\sigma$.

\subsection{Projection, orthogonality and geodesics}
\label{sec:BVD:geodesicbisector}

We start with an easy property of Bregman divergences.

\begin{property}[Three-point property] \label{prop:tpt}
For any triple $\v{p}, \v{q}$ and $\v{r}$ of points of $\set{X}$, we have:\\
$D_F(\v{p}||\v{q})+D_F(\v{q}||\v{r})=D_F(\v{p}||\v{r})+\dotproduct{\v{p}-\v{q}}{\v{r}'-\v{q}'}$.
\end{property}

The following lemma characterizes the {\em Bregman projection} of a
point onto a closed convex set $\mathcal{W}$.

\begin{lemma}[Bregman projection]
\label{lem-proj-unique}
For any $\pp$, there exists a unique point $\xx\in \mathcal{W}$ that
minimizes $D_F(\xx||\pp)$. We call this point the Bregman projection of
$\pp$ onto $\mathcal{W}$ and denote it $\pp_{\mathcal{W}}$.
\end{lemma}

\begin{proof}
If it is not the case, then define $\xx$ and $\yy$ two minimizers with
$D_F(\xx ||\pp) = D_F(\yy ||\pp) = l$. Since $\mathcal{W}$ is convex,
$(\xx + \yy)/2 \in {\mathcal{W}}$ and, since $D_F$ is strictly convex in its
first argument (see Section~\ref{sec:Bregman:Definition}), $D_F((\xx + \yy)/2 ||\pp) < D_F(\xx
||\pp)/2 + D_F(\yy ||\pp)/2$. But $D_F(\xx ||\pp)/2 + D_F(\yy
||\pp)/2=l$ yielding a contradiction.
\end{proof}

\begin{figure}
\centering
\includegraphics[bb=0 0 189 77, width=8cm]{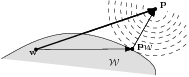}
\caption{Generalized Pythagoras' theorem for Bregman divergences: The projection $\v{p}_{\mathcal{W}}$ of point $\v{p}$ to a convex subset $\set{W}\subseteq\set{X}$.
For  convex subset $\set{W}$, we have $D_F(\v{w}||\v{p})\geq
D_F(\v{w}||\v{p}_{\mathcal{W}})+D_F(\v{p}_{\mathcal{W}}||\v{p})$ (with equality for and only for affine sets $\mathcal{W}$).  \label{fig:BregmanPytha}}
\end{figure}

We now introduce the notion of {\em Bregman orthogonality}. We say
that $\v{p}\v{q}$ is Bregman orthogonal to $\v{q}\v{r}$ iff
$D_F(\v{p}||\v{q})+D_F(\v{q}||\v{r})=D_F(\v{p}||\v{r})$ or
equivalently (by the Three-point property), if and only if 
$\dotproduct{\v{p}-\v{q}}{\v{r}'-\v{q}'}=0$. Observe the analogy with
Pythagoras' theorem in Euclidean space (see
Figure~\ref{fig:BregmanPytha}). Note also that the orthogonality
relation is not symmetric: the fact that $\v{p}\v{q}$ is Bregman
orthogonal to $\v{q}\v{r}$ does not necessarily imply that
$\v{q}\v{r}$ is Bregman orthogonal to $\v{p}\v{q}$.  More generally,
we say that $I \subseteq {\mathcal{X}}$ is \textit{Bregman orthogonal}
to $J \subseteq {\mathcal{X}}$ ($I \cap J \neq \emptyset$) iff for any
$\vector{p} \in I$ and $\vector{r} \in J$, there exists a $\vector{q}
\in I\cap J$ such that $\vector{p}\vector{q}$ is Bregman orthogonal to
$\v{q}\vector{r}$.

Notice that orthogonality is preserved in the gradient space. Indeed,
since $\dotproduct{\v{p}-\v{q}}{\v{r}'-\v{q}'}=
\dotproduct{\v{r}'-\v{q}'}{\v{p}-\v{q}}$, $\v{p}\v{q}$ is Bregman 
orthogonal to $\v{q}\v{r}$ iff $\v{r}'\v{q}'$ is Bregman orthogonal to
$\v{q}'\v{p'}$.

Let $\Gamma_F (\pp, \qq )$ be the image by $\gradIF$ of the line
segment $\pp '\qq '$, i.e.
\[ \Gamma_F (\pp, \qq ) = \{ \xx  \in \mathcal{X} : \xx '= (1-\lambda )\pp '+ \lambda \qq ', \lambda \in [0,1]\}.\]
By analogy, we rename the line segment $\v{p}\v{q}$ as
$$\Lambda (\v{p}, \v{q}) = \{\v{x}
\in {\set{X}}:  \v{x} =
(1 - \lambda) \v{p} + \lambda \v{q}, \lambda \in [0,1]\}$$ 

In the Euclidean case ($F(x)=\frac{1}{2} \| \v{x}\|^2$), $\Gamma_F
(\pp, \qq )=\Lambda (\v{p}, \v{q})$ is the unique geodesic path
joining $\v{p}$ to $\v{q}$ and it is orthogonal to the bisector
$H_F(\v{p},\v{q})$. For general Bregman divergences, we have similar
properties as shown next.

\begin{lemma}\label{cpq}
$\Gamma_F (\vector{p}, \vector{q})$ is Bregman orthogonal to the
Bregman bisector $H_F(\pp,\qq)$ while $\Lambda (\vector{p}, \vector{q})$
is Bregman orthogonal to $H_{F^*}(\pp,\qq)$.
\end{lemma}

\begin{proof}
Since $\pp$ and $\qq$ lie on different sides of $H_F(\pp,\qq)$,
$\Gamma_F (\vector{p}, \vector{q})$ must intersect $H_F(\pp,\qq)$.
Fix any distinct $\vector{x} \in \Gamma (\vector{p}, \vector{q})$
and
$\vector{y} \in H_F(\pp,\qq)$, and let
$\vector{t} \in \Gamma (\vector{p}, \vector{q})\cap H_F(\pp,\qq)$. To prove the first part of the lemma,
we need to show that $\dotproduct{\yy-\ttt}{\xx'-\ttt'}=0$.

Since $\ttt$ and $\xx$ both belong to $\in\Gamma_F (\vector{p},\vector{q})$, we have $\v{t}'-\xx ' = \lambda (\pp'-\qq')$, for some $\lambda\in\R$, and, since
$\yy$ and $\ttt$ belong to $H_F(\pp,\qq)$, we deduce from the equation
of $H_F(\pp,\qq)$ that $\dotproduct{\yy-\ttt}{\pp'-\qq'} =0$. We
conclude that $\dotproduct{\yy-\ttt}{\xx'-\ttt'} =0$, which proves
that $\Gamma_F (\vector{p}, \vector{q})$ is indeed Bregman orthogonal to 
$H_F(\pp,\qq)$.

The second part of the lemma is easily proved by using the fact that
orthogonality is preserved in the gradient space as noted above.
\end{proof}

\begin{figure}
\centering
\includegraphics[bb= 0 0 697 687, width=8cm]{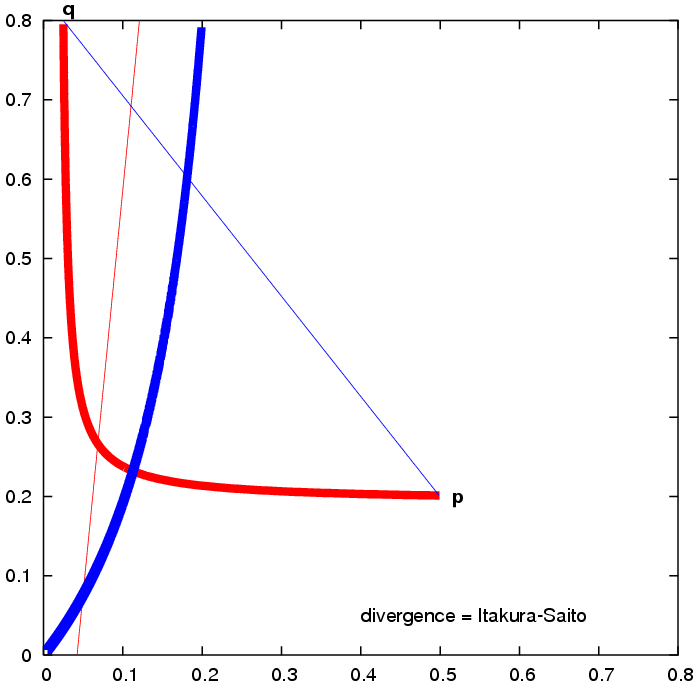}
\includegraphics[bb= 0 0 693 689, width=8cm]{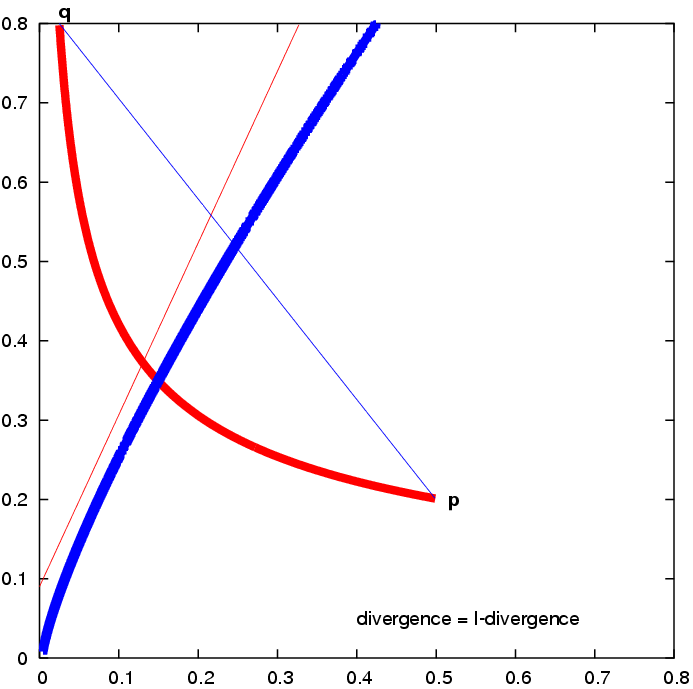}
\caption{\label{f-bb1}Bregman bisectors and their relationships with respect to 
$\Lambda (\vector{p}, \vector{q})$ (straight line segments) and $\Gamma_F (\vector{p},\vector{q})$ (bold curves), for the Itakura-Saito
divergence (left) and I-divergence (right). Bold curves become linear in $\set{X}'$;
colors depict the Bregman orthogonality relationships of Lemma
\ref{cpq}. }
\end{figure}

Figure \ref{f-bb1} shows Bregman bisectors and their relationships
with respect to $\Lambda (\vector{p}, \vector{q})$ and $\Gamma_F
(\vector{p},
\vector{q})$.


We now focus on characterizing Bregman geodesics.
First, recall that a parameterized curve ${\set{C}}$ between two points $\v{p}_0$ and $\v{p}_1$  is defined as a set ${\set{C}} = \{\v{p}_\lambda\}_{\lambda=0}^{1}$, which is continuous. In Riemannian geometry, geodesics are the curves that minimize the arc length with respect to the Riemannian metric~\cite{informationgeometry,anisotropicvor-2003}. 
Since embedding $\set{X}$ with a Bregman divergence does not yield a metric space, we define the following curve lengths:

\begin{eqnarray}
\ell_\Gamma({\mathcal{C}})  & = &  \int_{\lambda = 0}^{1} {D_F(\bm{p}_0 || \bm{p}_\lambda) \mathrm{d} \lambda}\:\:, \label{deflengthg}\\
\ell_\Lambda({\mathcal{C}}) & = & \int_{\lambda = 0}^{1} {D_F(\bm{p}_\lambda || \bm{p}_0) \mathrm{d} \lambda} \label{deflengthl}.
\end{eqnarray}

We now characterize the dual pair of geodesics and their lengths as follows:

\begin{lemma}\label{flem}
 Curve $\Gamma_F(\v{p}_0, \v{p}_1)$ (respectively straight line segment $\Lambda (\v{p}_0, \v{p}_1)$) minimizes $\int_{\lambda = 0}^{1} {D_F(\v{p}_0 || \v{p}_\lambda) \mathrm{d} \lambda}$ (respectively $\int_{\lambda = 0}^{1} {D_F(\bm{p}_\lambda || \bm{p}_0) \mathrm{d} \lambda}$) over all curves $\set{C} = \{\v{p}_\lambda\}_{\lambda=0}^{1}$. 
\end{lemma}

\begin{proof}
For any curve ${\set{C}}$ between $\v{p}_0$ and $\v{p}_1$, we measure the $\ell_\Gamma$ length as 
$\ell_\Gamma({\mathcal{C}})  =  \int_{\lambda} {D_F(\v{p}_\lambda ||\v{p}_0) \mathrm{d} \lambda}$.
Fix some inner point  $\v{p}\in\Gamma_F(\v{p}_0,\v{p}_1)\backslash\{\v{p}_0,\v{p}_1\}$. 
From the three-point property (Property~\ref{prop:tpt}), the set of points $\{\v{y}\in\set{X}\ |\ D_F(\v{y}||\v{p}_0)=D_F(\v{y}||\v{p})+D_F(\v{p}||\v{p}_0)\}$ is the hyperplane $H_{\v{p}}:  \dotproduct{\v{y}}{\v{h}}=b $ ($\v{h}$ is a perpendicular vector to the hyperplane)  which splits $\set{X}$ into two open half-spaces $H_{\v{p}}^+: \dotproduct{\v{y}}{\v{h}}>b$, and $H_{\v{p}}^-: \dotproduct{\v{y}}{\v{h}}<b$.  
Now, $H_{\v{p}}$ intersects $\Gamma(\v{p}_0,\v{p}_1)$ since $H_{\v{p}}$ separates $\v{p}_0$ from $\v{p}_1$.
Indeed, $H_{\v{p}}(\v{p}_0)=\dotproduct{\v{p}_0-\v{p}}{\v{p}_0'-\v{p}'}=D_F(\v{p}_0||\v{p})+D_F(\v{p}||\v{p}_0)>0$
and $H_{\v{p}}(\v{p}_1)=\dotproduct{\v{p}_1-\v{p}}{\v{p}_0'-\v{p}'}=\frac{\lambda-1}{\lambda} \dotproduct{\v{p}_1-\v{p}}{\v{p}_1'-\v{p}'}<0$ where $\v{p}'=\lambda\v{p}_0'+(1-\lambda)\v{p}_1'$ (with $\lambda\in ]0,1[$).
Therefore any connected path $\set{C}$ joining $\v{p}_0$ to $\v{p}_1$ has to intersect $H_{\v{p}}$.

To finish up, consider function $f: [0,1] \rightarrow {\mathcal{C}}$ with $f(0)=\v{p}_0$, $f(1)=\v{p}_1$, and $f(\lambda) \in {\set{C}} \cap H_{\v{p}_\lambda}$ otherwise, where it is understood that $\v{p}_\lambda$ is hereafter a point of $\Gamma_F (\v{p}_0, \v{p}_1)$.
Since $f(\lambda)\in H_{\v{p}(\lambda)}$, we have $D_F(f(\lambda)||\v{p}_0)=D_F(f(\lambda)||\v{p}_\lambda)+D_F(\v{p}_\lambda||\v{p}_0)\geq D_F(\v{p}_\lambda||\v{p}_0)$, with equality if and only if $f(\lambda)=\v{p}_\lambda$.
Thus we have
\begin{eqnarray*}
\ell_\Gamma (\Gamma_F (\v{p}_0, \v{p}_1)) = \int_{\lambda=0}^{1} {D_F(\v{p}_\lambda || \v{p}_0) \mathrm{d} \lambda} & \leq & \int_{\lambda=0}^{1} {D_F(f(\lambda) || \v{p}_0) \mathrm{d} \lambda} \leq \ell_\Gamma ({\set{C}})\:\:.
\end{eqnarray*} 

The case of $\Lambda (\v{p}_0, \v{p}_1)$ follows similarly from Legendre convex duality.

\end{proof}

\begin{corollary}
Since   $\Gamma_F(\v{p}_0, \v{p}_1)=\Gamma_F(\v{p}_1, \v{p}_0)$ (respectively, since $\Lambda (\v{p}_0, \v{p}_1)=\Lambda (\v{p}_1, \v{p}_0)$) we deduce that $\Gamma_F(\v{p}_0, \v{p}_1)$ minimizes also $\int_{\lambda = 0}^{1} {D_F(\v{p}_1 || \v{p}_\lambda) \mathrm{d} \lambda}$ (respectively, minimizes also  $\int_{\lambda = 0}^{1} {D_F(\v{p}_\lambda || \v{p}_1) \mathrm{d} \lambda}$) over all curves $\set{C} = \{\v{p}_\lambda\}_{\lambda=0}^{1}$.
\end{corollary}

Observe also that $\Gamma_F (\pp, \qq )$ is the unique geodesic
path joining $\pp$ to $\qq$ in $\set{X}$ for the metric
image by $\gradIF$ of the Euclidean metric.

Finally, we give a  characterization of these geodesics in information-theoretic spaces.
Recall that Banerjee et al.~\cite{j-cbd-2005} showed that Bregman divergences are in bijection with exponential families.
This was emphasized by  Theorem \ref{KLBD} that proved that the Kullback-Leibler divergence of probability density functions of the same exponential family $\set{E}_F$ is a Bregman divergence $D_F$ for the cumulant function $F$.
From this standpoint, $\Lambda (\v{p}, \v{q})$ and $\Gamma_F(\v{p}, \v{q})$  minimize the total Kullback-Leibler divergence, a characteristic that we choose to  call the  {\em information length} of a curve.
Since the Kullback-Leibler divergence is not symmetric, this justifies for the existence of two geodesics, one which appears to be linear when parameterized with
the natural affine coordinate system ($\thetav$), and the other that is linear in the expectation affine coordinate system ($\muv$). See also~\cite{informationgeometry}.

\begin{corollary}
Suppose $p(.|\bm{\theta}_0)$ and $p(.|\bm{\theta}_1)$ are probability density functions of the same exponential family ${\mathcal{E}}_F$. Then $\Gamma_F(\bm{\theta}_0, \bm{\theta}_1)$ (resp. $\Lambda (\bm{\theta}_0, \bm{\theta}_1)$) minimizes $\ell_\Gamma({\mathcal{C}})=\int_{\lambda=0}^1 \KL(\thetav_0||\thetav_\lambda) \mathrm{d}\lambda$ (resp. $\ell_\Lambda({\mathcal{C}})=\int_{\lambda=0}^1 \KL(\thetav_\lambda||\thetav_0) \mathrm{d}\lambda$) over all curves $\set{C} = \{p(.|\bm{\theta}_\lambda)\}_{\lambda=0}^{1}$.
\end{corollary}

\section{Bregman Voronoi diagrams \label{sec:BVD}}

Let $\set{S}=\{\v{p}_1, ..., \v{p}_n\}$ be a finite point set of
$\set{X} \subset \mathbb{R}^{d}$.  To each point $\v{p}_i$ is attached
a $d$-variate  continuous function $D_i$ defined over
$\set{X}$. We define the {\em lower envelope} of the functions as the
graph of $\min_{1\leq i\leq n} D_i$ and their {\em minimization
diagram} as the subdivision of $\set{X}$ into cells such that, in each
cell, $\arg\min_i f_i$ is fixed.

The Euclidean Voronoi diagram is the minimization diagram for
$D_i(\xx)= \| \v{x}-\pp _i\| ^2$.  In this section, we introduce Bregman
Voronoi diagrams as minimization diagrams of Bregman divergences
(see Figure~\ref{fig:mindiagrams}).

We define three types of  Bregman Voronoi diagrams in
\S\ref{sec:3typesBVD}. We establish a correspondence between 
Bregman Voronoi diagrams and polytopes in
\S\ref{sec:Polytopes} and with power diagrams in \S\ref{sec:PowerDiagram}.
These correspondences lead to tight combinatorial bounds and efficient
algorithms. Finally, in \S\ref{sec:generalizedBVD}, we give two
generalizations of Bregman Voronoi diagrams; $k$-order and $k$-bag diagrams.

We note $\set{S}'=\{\grad_F(\v{p}_i), i=1,\ldots ,n\}$ the {\it
gradient point set} associated to $\set{S}$.

\subsection{Three types of diagrams}\label{sec:3typesBVD}

Because Bregman divergences are not necessarily symmetric, we
associate to each site $\v{p}_i$ {\it two types} of distance
functions, namely $D_i(\v{x})=D_F(\v{x}||\v{p}_i)$ and
$D_i'(\v{x})=D_F(\v{p}_i||\v{x})$.  The {\it minimization diagram} of
the $D_i$, $i=1,\ldots ,n$, is called the {\em first-type} Bregman
Voronoi diagram of $\set{S}$, which we denote by
$\vor_{F}(\set{S})$. The $d$-dimensional cells of this diagram are in
{\em 1-1 correspondence} with the sites $\v{p}_i$ and the $d$-dimensional cell
of $\v{p}_i$ is defined as $$
\vor_F(\v{p}_i)\equaldef \{\v{x}\in\set{X}\ |\ D_F(\v{x}||\v{p}_i) \leq D_F(\v{x}||\v{p}_j) \ \forall\v{p}_j\in\set{S}. \}
$$

Since the Bregman bisectors of the first-type are hyperplanes, the
cells of any diagram of the first-type are convex
polyhedra. Therefore,  first-type Bregman Voronoi diagrams are {\em
affine} diagrams~\cite{affinevoronoi-1987,ak-vd-00}.  

Similarly, the minimization diagram of the $D_i'$, $i=1,\ldots ,n$, is
called the {\em second-type} Bregman Voronoi diagram of $\set{S}$,
which we denote by $\vor'_{F}(\set{S})$.  A cell in
$\vor'_{F}(\set{S})$ is associated to each site $\v{p}_i$ and is
defined as above with permuted divergence arguments: $$
\vor'_F(\v{p}_i)\equaldef\{\v{x}\in\set{X}\ |\ D_F(\v{p}_i||\v{x})
\leq D_F(\v{p}_j||\v{x}) \
\forall\v{p}_j\in\set{S}. \}
$$

In contrast with the diagrams of the first-type, the diagrams of the
second type have, in general, curved faces.

Figure~\ref{fig:dbv} illustrates these Bregman Voronoi diagrams for
the Kullback-Leibler and the Itakura-Saito divergences. Note that
the Euclidean Voronoi diagram is a Bregman Voronoi diagram since
$\vor(\set{S})=\vor_{F}(\set{S})=\vor'_F(\set{S})$ for $F(\vector{x})=
\| \xx \| ^2$.

For asymmetric Bregman divergences $D_F$, we can further consider the
symmetrized Bregman divergence $S_F=D_{\tilde F}$ and define a {\em
third-type} Bregman Voronoi diagram $\vor''_{F}(\set{S})$.  The
cell of $\vor''_{F}(\set{S})$ associated to  site $\pp _i$ is  defined as: $$
\vor''_F(\v{p}_i)\equaldef \{\v{x}\in\set{X}\ |\ S_F(\v{x},\v{p}_i) \leq S_F(\v{x},\v{p}_j) \ \forall\v{p}_j\in\set{S}. \}
$$

\def\ttt{7.5cm}
\begin{figure}
\centering
\begin{tabular}{cc}
\fbox{\includegraphics[bb=0 0 1024 1024,width=\ttt]{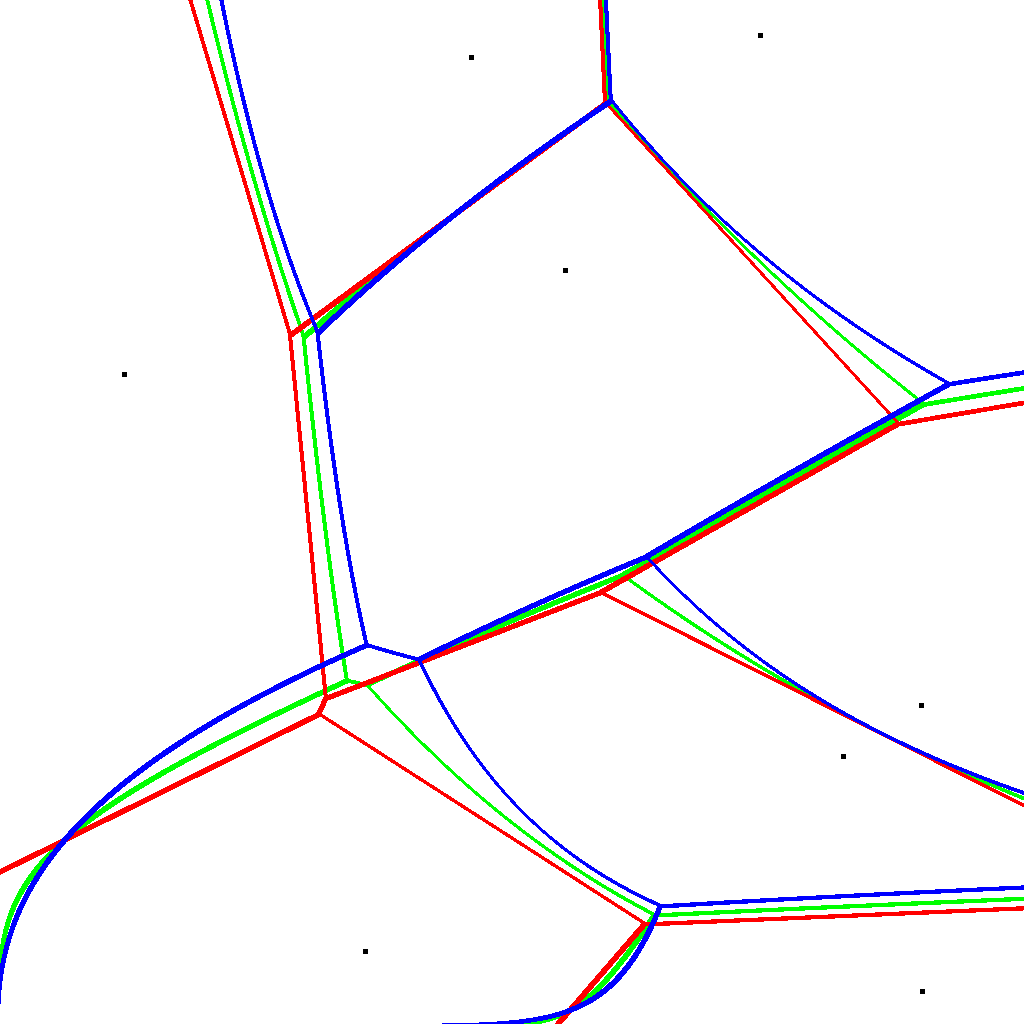}} &
\fbox{\includegraphics[bb=0 0 1024 1024,width=\ttt]{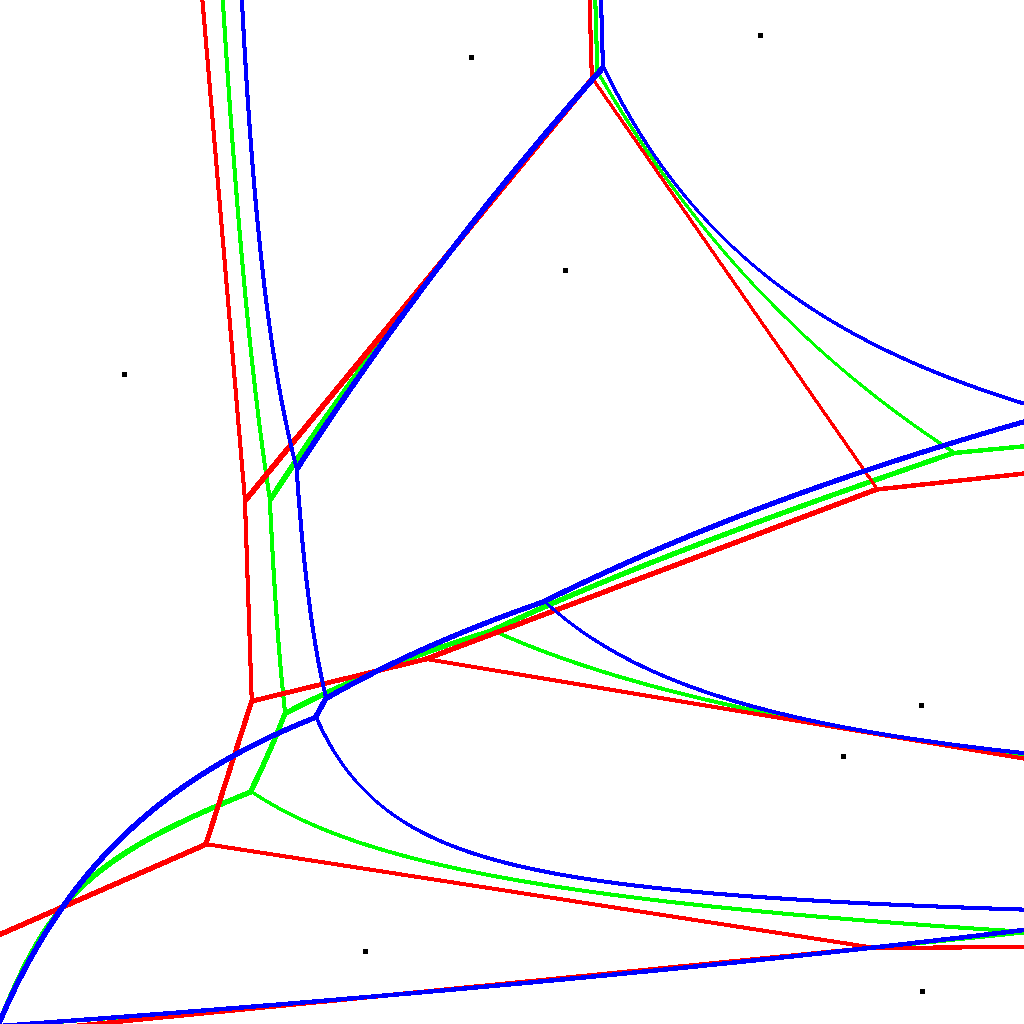}}\\
(a) & (b)
\end{tabular}
\caption{\label{fig:dbv}Three types of Bregman Voronoi diagrams for (a) the Kullback-Leibler and (b) the Itakura-Saito divergences. 
First-type affine Bregman Voronoi diagram (red), second-type Bregman Voronoi diagram (blue) and symmetrized Bregman Voronoi diagram (green).
}
\end{figure}

From the Legendre duality between divergences, we deduce
correspondences between the diagrams of the first and the second
types. As usual, $F^*$ is the convex conjugate of $F$.

\begin{lemma}
\label{th-bregman12}
$\vor'_F(\set{S})= \gradIF(\vor_{F^*}(\set{S}'))$ and
$\vor_F(\set{S}) = \gradIF(\vor'_{F^*}(\set{S}'))$.
\end{lemma}

\begin{proof}
By Lemma~\ref{lem-dualdiv}, we have
$D_F(\v{x}||\v{y})=D_{F^*}(\v{y}'||\v{x}')$,
which gives $\vor_F(\vector{p}_i)=\{\vector{x}\in\set{X}\ |\
D_{F^*}(\vector{p}_i'||\vector{x}') \leq
D_{F^*}(\vector{p}_j'||\vector{x}') \ \forall\v{p}_j'\in\set{S}'
\}=\gradIF (\vor_{F^*}'(\v{p}_i')).$
The proof of the second part follows the same path.
\end{proof}

Hence, constructing the second-type curved diagram
$\vor'_{F}(\set{S})$ reduces to constructing an affine diagram in the
gradient space $\set{X}'$ (and map the cells by $\gradF^{-1}$).

Let us end this section by considering the case of symmetrized Bregman
divergences introduced in \S\ref{sec:Bregman:SymmetrizedDivergence}:
$S_F(\v{p},\v{q})=D_{\tilde F}(\tilde\v{p}||\tilde\v{q})=D_{\tilde
F}(\tilde\v{q}||\tilde\v{p})$ where $\tilde F$ is a $2d$-variate
function and $\tilde\xx = [\v{x}\ \v{x}']^T$. As already noted in
\S\ref{sec:Bregman:SymmetrizedDivergence}, $\tilde\xx $ lies on the 
$d$-manifold  $\tilde\set{X}=\{ [\v{x}\
\v{x}']^T \ |\
\v{x}\in\mathbb{R}^d\}$. It follows that the symmetrized Voronoi diagram
$\vor''_{F}(\set{S})$ is the projection of the restriction to
$\tilde\set{X}$ of the affine diagram $\vor_{\tilde F}(\tilde\set{S})$
of $\R ^{2d}$ where $\tilde\set{S}= \{ \tilde{\pp}_i, \pp_i\in
\set{S}\}$.  Hence, computing the symmetrized Voronoi diagram of
$\set{S}$ reduces to: 
\begin{enumerate}
\item  computing the first-type Bregman Voronoi
diagram $\vor_{\tilde F}(\tilde\set{S})$ of $\R^{2d}$, 
\item intersecting
the cells of this diagram with the manifold $\tilde\set{X}$, and
\item
 projecting all points of $\vor_{\tilde
F}(\tilde\set{S})\cap\tilde\set{X}$ to $\set{X}$ by simply dropping
the last $d$ coordinates.
\end{enumerate}

\subsection{Bregman Voronoi diagrams from polytopes\label{sec:Polytopes}}

\begin{figure}
\centering
\begin{tabular}{|cc|}\hline
\multicolumn{2}{|c|}{Squared Euclidean distance}\\ \hline\hline
\includegraphics[bb=0 0 1020 1019, width=7.5cm]{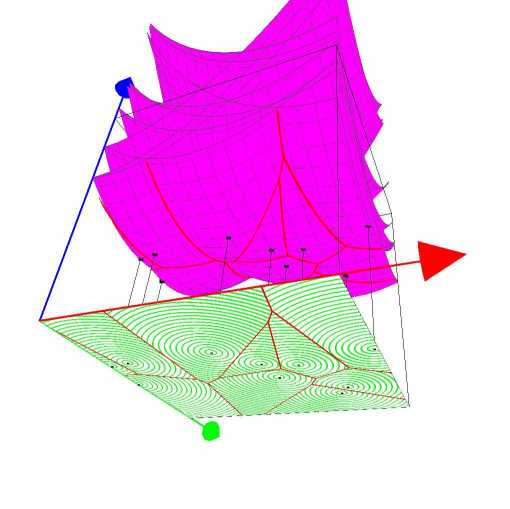} &
\includegraphics[bb=0 0 1018 1020, width=7.5cm]{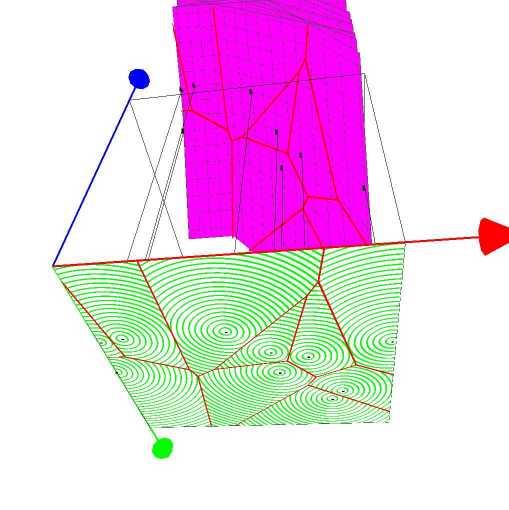}
\\
(a) & (b)\\ \hline
\multicolumn{2}{|c|}{Kullback-Leibler divergence}\\ \hline\hline
\includegraphics[bb=0 0 1006 999, width=7.5cm]{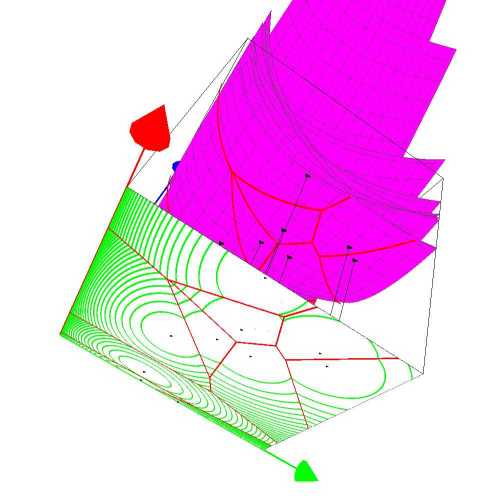}
&
\includegraphics[bb=0 0 1017 106, width=7.5cm]{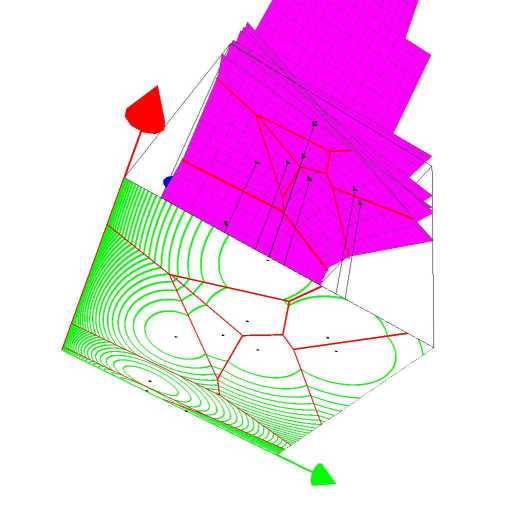}\\
(c) & (d) \\ \hline
\end{tabular}
\caption{Voronoi diagrams as minimization diagrams.
 The first row shows minimization diagrams for the Euclidean distance
 and the second row shows minimization diagrams for the
 Kullback-Leibler divergence. In the first column, the functions are
 the non-linear functions $D_i(\xx )$ and, in the second column, the
 functions are the linear functions $H_{\pp_i}(\xx)$, both leading to the same minimization diagrams. Isolines are
 shown in green.\label{fig:mindiagrams}}
\end{figure}

Let $H_{\v{p}_i}$, $i=1,\ldots ,n$, denote the hyperplanes of
$\hat\set{X}$ defined in
\S\ref{sec:SpherePolarity}.  For any $\xx\in \set{X}$, we have following Lemma~\ref{lem-code-BD}
\[ D_F(\xx||\pp_i) \leq D_F(\xx||\pp_j) \Longleftrightarrow
H_{\pp_i}(\xx) \geq H_{\pp_j}(\xx).\] The first-type Bregman Voronoi
diagram of $\set{S}$ is therefore the maximization diagram of the $n$
linear functions $H_{\pp_i}(\xx)$  whose graphs are the
hyperplanes $H_{\pp_i}$ (see
Figure~\ref{fig:mindiagrams}). Equivalently, we have

\begin{theorem}
\label{th-polytope}
The first-type Bregman Voronoi diagram $\vor_F(\mathcal{S})$ is
obtained by projecting by $\proj_{\set{X}}$ the faces of the
$(d+1)$-dimensional convex polyhedron  $\mathcal{H}=\cap_{i}
H_{\v{p}_i}^{\uparrow}$ of $\mathcal{X}^+$ onto $\mathcal{X}$.
\end{theorem}

From McMullen's upperbound theorem~\cite{McMullen1971}
and Chazelle's optimal half-space intersection algorithm~\cite{Chazelle1993}, we know that  the intersection of $n$ halfspaces of $\R ^d$ has
complexity $\Theta(n^{
\floor{ \frac{d}{2} } })$ and can be computed in optimal-time
$\Theta(n\log n+n^{ \floor{ \frac{d}{2} } })$ for any fixed
dimension $d$. From Theorem \ref{th-polytope} and Lemma
\ref{th-bregman12}, we then deduce the following theorem.

\begin{theorem}
\label{th-complexity}
The Bregman Voronoi diagrams of type 1 or 2 of a set of $n$
$d$-dimensional points have complexity $\Theta(n^{ \floor{
\frac{d+1}{2} } })$ and can be computed in optimal time $\Theta(n\log
n+n^{ \floor{ \frac{d+1}{2} } })$.
The third-type Bregman Voronoi diagram for the symmetrized Bregman divergence of a set of $n$ $d$-dimensional points has complexity $O(n^d)$
and can be obtained in  time $O(n^d)$.
\end{theorem}

Apart from Chazelle's algorithm, several other algorithms are known
for constructing the intersection of a finite number of halfplanes,
especially in the 2- and 3-dimensional cases. See \cite{compgeom-1998,ak-vd-00} for further
references.

\subsection{Bregman Voronoi diagrams from power diagrams\label{sec:PowerDiagram}}

The power distance of a point $\vector{x}$ to a Euclidean ball
$B=\ballE(\vector{p},r)$ is defined as
$||\vector{p}-\vector{x}||^2-r^2$. Given $n$ balls $B_i=\ballE
(\v{p}_i, r_i)$, $i=1,\ldots ,n$, the {\em power diagram} (or Laguerre diagram) of the $B_i$ is
defined as the minimization diagram of the corresponding $n$ functions
$D_i(\xx )=||\vector{p}_i-\vector{x}||^2-r^2$. The power bisector of
any two balls $\ballE(\v{p}_i,r_i)$ and $\ballE(\v{p}_j,r_j)$ is the radical
hyperplane of equation
$2\dotproduct{\v{x}}{\v{p}_j-\v{p}_i}+||\v{p}_i||^2-||\v{q}_j||^2+r_j^2-r_i^2=0$.
Thus power diagrams are affine diagrams. In fact, as shown by
Aurenhammer~\cite{powerdiagrams-1987,compgeom-1998}, {\em any} affine
diagram is {\em identical} to the power diagram of a set of corresponding balls.
In general, some balls may have an empty cell in their power
diagram.

\begin{figure}
\centering
\begin{tabular}{|c|c|} \hline
Affine Bregman Voronoi diagram & Equivalent Power diagram\\ \hline\hline
\includegraphics[bb=0 0 472 465, width=5.5cm]{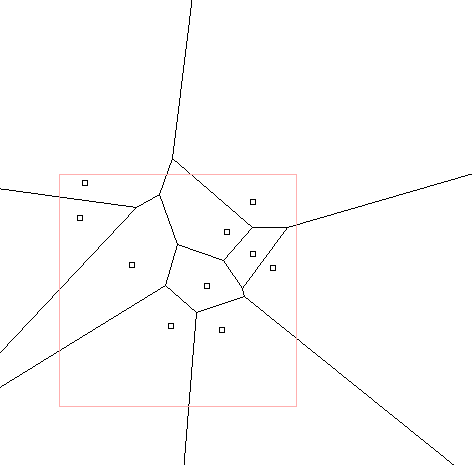}&
\includegraphics[bb=0 0 472 465, width=5.5cm]{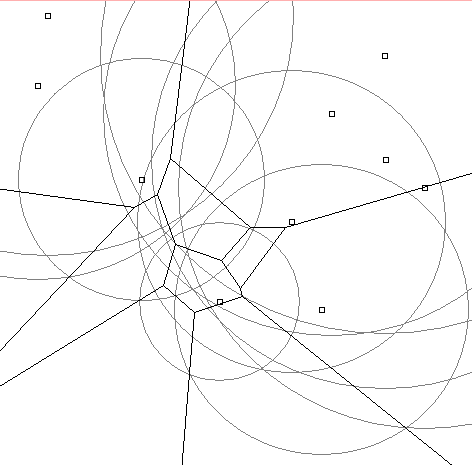} \\ \hline
\multicolumn{2}{|c|}{(a) Squared Euclidean distance ($F(\v{x})=||\v{x}||^2$)}\\ \hline\hline
\includegraphics[bb=0 0 472 467, width=5.5cm]{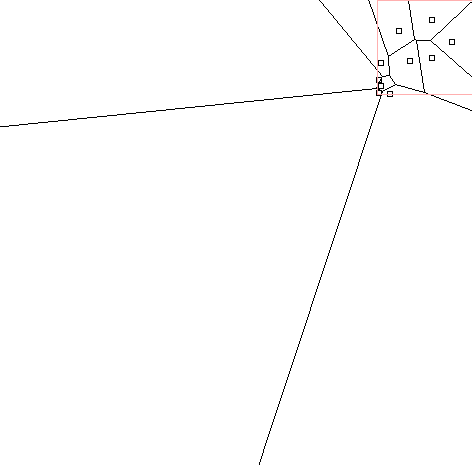} &
\includegraphics[bb=0 0 471 467, width=5.5cm]{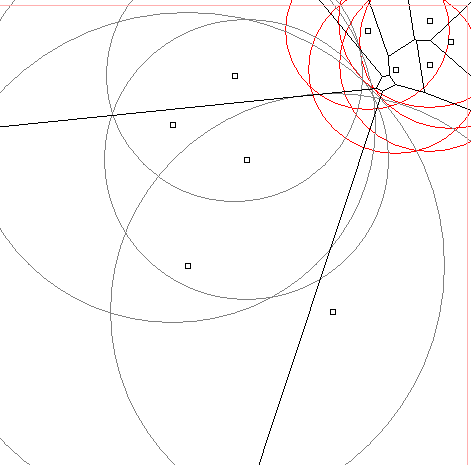} \\ \hline
\multicolumn{2}{|c|}{(b) Kullback-Leibler divergence ($F(\v{x})=\sum_i x_i\log x_i$)}\\ \hline\hline
\includegraphics[bb=0 0 473 467, width=5.5cm]{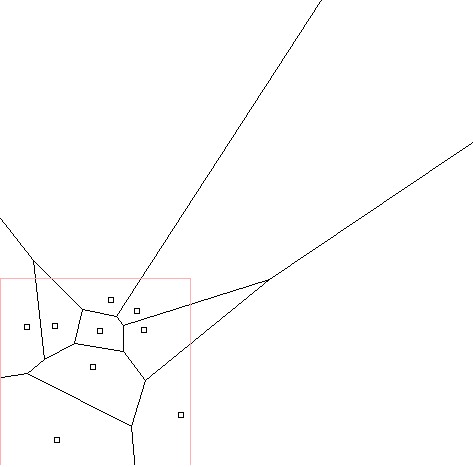} &
\includegraphics[bb=0 0 472 467, width=5.5cm]{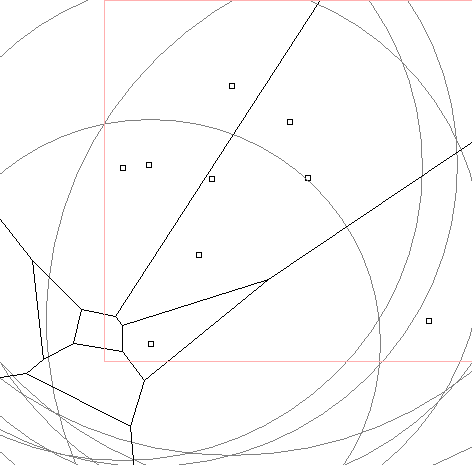} \\ \hline
\multicolumn{2}{|c|}{(c) Exponential loss divergence ($F(\v{x})=\sum_i \exp x_i$)}\\ \hline
\end{tabular}
\caption{Affine Bregman Voronoi diagrams (left column) can be computed as power diagrams (right column). 
Illustrations for the squared Euclidean distance~(a), Kullback-Leibler divergence~(b), and exponential divergence~(c). Circles are drawn either in grey to denote positive radii, or in red to emphasize imaginary radii. Observe that although some cells of the power diagrams may be empty, all cells of the affine Bregman Voronoi diagram are necessarily non-empty.
\label{fig:bvd2pd}}
\end{figure}

Since Bregman Voronoi diagrams of the first type are affine diagrams,
Bregman Voronoi diagrams are power
diagrams~\cite{powerdiagrams-1987,compgeom-1998} in disguise. The
following theorem makes precise the correspondence between Bregman 
Voronoi diagrams and power diagrams (see Figure~\ref{fig:bvd2pd}).

\begin{theorem}
The first-type Bregman Voronoi diagram of $n$ sites
is identical to the power diagram of the $n$ Euclidean spheres 
of equations
$$\dotproduct{\v{x}-\v{p}_i'}{\v{x}-\v{p}_i'} = \dotproduct{\vector{p}_i'}{\vector{p}_i'}+
2(F(\vector{p}_i) - \dotproduct{\vector{p}_i}{\vector{p}_i'}), \; i=1,\ldots ,n. $$
\label{th-power}
\end{theorem}

\begin{proof}
We have
\begin{eqnarray*}
&& D_F (\v{x}||\v{p}_i)   \leq  D_F (\v{x}||\v{p}_j)\\
&& \Longleftrightarrow   -F(\v{p}_i) -
\dotproduct{\v{x}-\vector{p}_i}{\vector{p}_i'} \leq - F(\v{p}_j) -
\dotproduct{\v{x}-\vector{p}_j}{\vector{p}_j'}
\end{eqnarray*}

\noindent Multiplying twice the last inequality, and adding $\dotproduct{\v{x}}{\v{x}}$ to both  sides yields
\begin{eqnarray*}
&&\dotproduct{\v{x}}{\v{x}} - 2 \dotproduct{\v{x}}{\vector{p}_i'} -2
F(\v{p}_i) +2
\dotproduct{\v{p}_i}{\v{p}'_i}
\leq \dotproduct{\v{x}}{\v{x}} - 2 \dotproduct{\v{x}}{\vector{p}_j'} -2 F(\v{p}_j) +2
\dotproduct{\v{p}_j}{\v{p}'_j}\\
&&\Longleftrightarrow   
\dotproduct{\v{x}-\v{p}'_i}{\v{x}-\v{p}'_i}- r_i^2\leq \dotproduct{\v{x}-\v{p}'_j}{\v{x}-\v{p}'_j}- r_j^2,
\end{eqnarray*}
where $r_i^2= \dotproduct{\vector{p}_i'}{\vector{p}_i'}+
2(F(\vector{p}_i) - \dotproduct{\vector{p}_i}{\vector{p}_i'})$ and
$r_j^2= \dotproduct{\vector{p}_j'}{\vector{p}_j'}+ 2(F(\vector{p}_j) -
\dotproduct{\vector{p}_j}{\vector{p}_j'})$.  The last inequality means
that the power of $\v{x}$ with respect to the Euclidean (possibly
imaginary) ball $B(\v{p}'_i,r_i)$ is no more than the power of
$\v{x}$ with respect to the Euclidean (possibly imaginary) ball
$B(\v{p}'_j,r_j)$.
\end{proof}

As already noted, for $F(\vector{x})=\frac{1}{2}\| \xx \| ^2$,
$\vor_F(\mathcal{S})$ is the Euclidean Voronoi diagram of
$\mathcal{S}$.  Accordingly, the theorem says that the centers of the
spheres are the $\pp_i$ and $r_i^2=0$ since $\pp_i'=\pp_i$.
Figure~\ref{fig:bvd2pd} displays  affine Bregman Voronoi diagrams\footnote{See Java\texttrademark{} applet at \url{http://www.csl.sony.co.jp/person/nielsen/BVDapplet/}} and their equivalent power diagrams for the squared Euclidean, Kullback-Leibler and exponential divergences.

Note that although the affine Bregman Voronoi diagram obtained by scaling the divergence $D_F$ by a factor $\lambda>0$ does not change, the equivalent power diagrams
 are not {\em strictus senso} identical since the centers of corresponding Euclidean balls and radii are mapped differently. See the example of the squared Euclidean distance depicted in Figure~\ref{fig:bvd2pd}(a). Since Power diagrams are well defined ``everywhere'', this equivalence relationship provides a natural way to extend the scope of definition of Bregman Voronoi diagrams from $\set{X}\subset\mathbb{R}^d$ to the full space $\mathbb{R}^d$. 
 (That is, Bregman Voronoi diagrams are power diagrams restricted to $\set{X}$.)

To check that associated balls may be potentially  imaginary, consider for example, the Kullback-Leibler divergence.
The Bregman generator function is $F(\v{x})=\sum_{i} x_i\log x_i$ and the gradient is $\gradF(\v{x})=[\log x_1\ ...\ \log x_d]^T$.
A point $\v{p}=[p_1\ ...\ p_d]^T\in\set{X}$ maps to a Euclidean ball of center $\v{p}'=[\log p_1\ ...\ \log p_d]^T$ with radius 
$r_{\v{p}}^2=\sum_{i} (\log^2 p_i -2p_i)$.  
Thus for  points $\v{p}$ with  coordinates $p_i>\frac{1}{2}\log p_i^2$ for $i\in\{1, ..., d\}$, the squared radius $r_{\v{p}}^2$ is negative, yielding an imaginary ball. See Figure~\ref{fig:bvd2pd}(b).

It is also to be observed that not all power diagrams are Bregman
Voronoi diagrams. Indeed, in power diagrams, some balls may have  
empty cells while each site has necessarily a non empty cell in a Bregman Voronoi
diagram (See Figure~\ref{fig:bvd2pd} and Section
\ref{sec-weighted} for a further discussion at this point).

Since there exist fast algorithms for constructing power
diagrams~\cite{cgal:pt-tds3-06}, Theorem~\ref{th-power} provides an
{\em efficient} way to construct Bregman Voronoi diagrams.

\subsection{Generalized Bregman divergences and their Voronoi diagrams}
\label{sec:generalizedBVD}

\subsubsection*{Weighted Bregman Voronoi diagrams\label{sec-weighted}}

Let us associate to each site $\v{p}_i$ a weight $w_i\in \R$. We
define the {\em weighted divergence} between two weighted points as
$W\!D_F(\v{p}_i||\v{p}_j) \equaldef D_F(\v{p}_i||\v{p}_j) + w_i-w_j$.
We can define bisectors and weighted Bregman Voronoi diagrams in very
much the same way as for non weighted divergences. 
The Bregman Voronoi region associated to the weighted point 
$(\vector{p}_i,w_i)$ is defined as
$$\vor_F(\vector{p}_i,w_i)= \{\vector{x}\in\mathcal{X}\ |\
D_F(\vector{x}||\vector{p}_i)+w_i \leq D_F(\vector{x}||\vector{p}_j)+w_j
\ \forall\vector{p}_j\in\mathcal{S} \}.$$ 
Observe that the bisectors of the first-type diagrams are still
hyperplanes and that the diagram can be obtained as the projection of
a convex polyhedron or as the power diagram of a finite set of
balls. The only difference with respect to the construction of
Section~\ref{sec:Polytopes} is the fact that now the hyperplanes
$H_{\v{p}_i}$ are no longer tangent to $\mathcal{F}$ since they are
{\it shifted} by a $z$-displacement of length $w_i$. Hence
Theorem~\ref{th-complexity} extends to weighted Bregman Voronoi
diagrams. 

\begin{theorem}
The weighted Bregman Voronoi diagrams of type 1 or 2 of a set of $n$
$d$-dimensional points have complexity $\Theta(n^{ \floor{
\frac{d+1}{2} } })$ and can be computed in optimal time $\Theta(n\log
n+n^{ \floor{ \frac{d+1}{2} } })$.
\label{th-WBVD}
\end{theorem}

\subsubsection*{$k$-order Bregman Voronoi diagrams}

We define the $k$-order Bregman Voronoi diagram of $n$ punctual sites
of $\mathcal{X}$ as the subdivision of $\mathcal{X}$ into cells such
that each cell is associated to a subset $\set{T}\subset \set{S}$ of
$k$ sites and consists of the points of $\mathcal{X}$ whose divergence
to any site in $\set{T}$ is less than the divergence to the sites not
in $\set{T}$.  Similarly to the case of higher-order Euclidean Voronoi
diagrams, we have:

\begin{theorem}
The $k$-order Bregman Voronoi diagram of $n$ $d$-dimensional points is
a weighted Bregman Voronoi diagram.
\end{theorem}

\begin{proof}
Let $\set{S}_1,\set{S}_2,\ldots$ denote the  subsets of $k$ points of $\set{S}$ and write
\begin{eqnarray*}
D_i(\v{x}) & =& \frac{1}{k}\; \sum_{\pp_j\in \set{S}_i} D_F(\v{x}||\v{p}_j)\\
&=&  F(\v{x}) -\frac{1}{k}\;\sum_{\pp_j\in \set{S}_i}F(\v{p}_j) 
+  \frac{1}{k}\; \sum_{\pp_j\in \set{S}_i}  \dotproduct{\v{x}-\v{p}_j}{\v{p}_j'}\\
&=& F(\v{x}) -F(\v{c}_i) - \dotproduct{\v{x}-\v{c}_i}{\v{c}_i'} + w_i\\
& =& WD_F(\v{x}||\v{c}_i)
\end{eqnarray*}
where $\v{c}_i= \gradIF\left( \frac{1}{k}\;\sum_{j\in
S_i}\v{p}_j'\right) $ and the weight associated  to $\v{c}_i$ is $w_i= F(\v{c}_i) -
\dotproduct{\v{c}_i}{\v{c}_i'} -\frac{1}{k}\;\sum_{j\in
S_i} \left( F(\v{p}_j) +
\dotproduct{\v{p}_j}{\v{p}_j'} \right)$.

Hence, $\set{S}_i$ is the set of the $k$ nearest neighbors of $\v{x}$  iff
$D_i(\v{x}) \leq D_j(\v{x})$ for all $j$ or,
equivalently, iff $\v{x}$ belongs to the cell of $\v{c}_i$ in the weighted Bregman
Voronoi diagram of the $\v{c}_i$.
\end{proof}

\subsubsection*{$k$-bag Bregman Voronoi diagrams\label{sec:kBVD}}

\def\bnabla{{\bm\nabla}}
\def\balpha{{\bm\alpha}}
\def\m#1{\mathbf{#1}}
\def\p{{\v{p}}}
\def\q{{\v{q}}}

Let $F_1, ..., F_k$ be $k$ strictly convex and differentiable functions, and $\balpha=[\alpha_1\ ...\ \alpha_k]^T\in\mathbb{R}^k_+$ a vector of positive weights.
Consider the $d$-variate function $F_\balpha=\sum_{l=1}^k \alpha_l F_l$.
By virtue of the positive additivity property rule of Bregman basis functions (Property~\ref{prop:linearoperator}), $D_{F_{\balpha}}$ is a Bregman divergence.

Now consider a set $\set{S}=\{\v{p}_1, ..., \v{p}_n\}$ of $n$ points of $\mathbb{R}^d$. 
To each site $\v{p}_i$, we associate a weight vector $\balpha_i=[\alpha_i^{(1)}\ ...\ \alpha_i^{(k)}]^T$ inducing a Bregman divergence $D_{F_{\balpha_i}}(\v{x}||\v{p}_i)\equaldef D_{\balpha_i}(\v{x}||\v{p}_i)$ anchored at that site. 
Let us consider the first-type of $k$-bag Bregman Voronoi diagram ($k$-bag BVD for short).
The first-type bisector $K_F(\v{p}_i,\v{p}_j)$ of two weighted points $(\v{p}_i,\balpha_i)$ and $(\v{p}_j,\balpha_j)$ is the locus of points $\v{x}$ at equidivergence to $\v{p}_i$ and $\v{p}_j$.
That is,  $K_F(\v{p}_i,\v{p}_j)=\{\v{x}\in\set{X}\ |\ D_{\balpha_i}(\v{x}||\v{p}_i)  = D_{\balpha_j}(\v{x}||\v{p}_j)\}$. The equation of the bisector is simply obtained using the definition of Bregman divergences (Eq.~\ref{eq:basicdf}) as  

$$
F_{\balpha_i}(\v{x})-F_{\balpha_i}(\v{p}_i)-\dotproduct{\v{x}-\v{p}_i}{\bnabla F_{\balpha_i}(\v{p}_i)}  =  F_{\balpha_j}(\v{x})-F_{\balpha_j}(\v{p}_j)-\dotproduct{\v{x}-\v{p}_j}{\bnabla F_{\balpha_i}(\v{p}_j)}.
$$

This yields the equation of the first-type bisector $K_F(\v{p}_i,\v{p}_j)$
\begin{equation}\label{eq:kbvd}
\sum_{l=1}^k (\balpha_{i}^{(l)}-\balpha_{j}^{(l)})F_l(\v{x}) - \dotproduct{\v{x}}{\bnabla F_{\balpha_j}(\v{p}_j)-\bnabla F_{\balpha_i}(\v{p}_i) } + c=0,\\
\end{equation}
\noindent where $c$ is a {\em constant} depending on weighted sites $(\v{p}_i,\balpha_i)$ and $(\v{p}_j,\balpha_j)$.
Note that the equation of the first-type $k$-bag BVD bisector is linear if and only if  $\balpha_i=\balpha_j$ (i.e., the case of standard BVDs).

Let us consider the linearization lifting $\v{x}\mapsto \hat\v{x}=[\v{x}\ F_1(\v{x})\ ...\ F_k(\v{x})]^T$ that maps a point $\v{x}\in\mathbb{R}^d$ into a point $\hat{\v{x}}$ in $\mathbb{R}^{d+k}$.  Then Eq.~\ref{eq:kbvd}  becomes linear, namely $\dotproduct{\hat\v{x}}{\v{a}}+c=0$ with 

$$
\v{a}=\left[
\begin{array}{c}
\bnabla F_{\balpha_j}(\v{p}_j)-\bnabla F_{\balpha_i}(\p_i)\\
\balpha_i-\balpha_j\\
\end{array}
\right]  \in\mathbb{R}^{d+k}.
$$

That is,  first-type bisectors of a $k$-bag BVD are  hyperplanes of $\mathbb{R}^{d+k}$.
Therefore the complexity of a $k$-bag Voronoi diagram is  at most $O(n^{\floor{\frac{k+d}{2}}})$, since it can be obtained 
as the intersection of the affine Voronoi diagram in $\mathbb{R}^{d+k}$ with the convex $d$-dimensional submanifold  $\{\hat\v{x}=[\v{x}\ F_1(\v{x})\ ...\ F_k(\v{x})]^T\ |\ \v{x}\in\mathbb{R}^d\}$.

\begin{theorem}
The $k$-bag Voronoi diagram (for $k>1$) on a bag of $d$-variate Bregman divergences of a set of $n$ points of $\mathbb{R}^d$ has
combinatorial complexity  $O(n^{\floor{\frac{k+d}{2}}})$ and can be computed within the same time bound.
\end{theorem}

Further, using the Legendre transform, we define a second-type (dual) $k$-bag BVD.
We have $\bnabla F_{\balpha}=\sum_{l=1}^k \alpha_l\bnabla F_l$ and $ F_{\balpha}^*=\int \bnabla F_{\balpha}^{-1}$. (Observe that $F_{\balpha}^*\not =\sum_{l=1}^k \alpha_lF_l^*$ in general.)

$k$-bag Bregman Voronoi diagrams are closely related to the  anisotropic diagrams of Labelle and Shewchuk~\cite{anisotropicvor-2003} that associate to {\em each} point $\v{x}\in\set{X}$ a metric tensor $\m{M}_{\v{x}}$ which tells how lengths and angles should be measured from the local perspective of $\v{x}$.
Labelle and Shewchuk relies on a deformation tensor (ideally defined everywhere) to compute the distance between any two points $\v{p}$ and $\v{q}$ from the perspective of $\v{x}$ as $d_{\v{x}}(\v{p},\v{q})=\sqrt{(\v{p}-\v{q})^T\m{M}_{\v{x}}(\v{p}-\v{q})}$. 
Let $d_{\v{x}}(\v{p})=d_{\v{x}}(\v{x},\v{p})$.
The  anisotropic Voronoi diagram, which approximates the ideal but computationally prohibitive Riemannian Voronoi diagram, is defined as the arrangement of the following anisotropic Voronoi cells:

$$
\Vor(\v{p}_i)=\{\v{x}\in\set{X}\ |\ d_{\v{p}_i}(\v{x})\leq d_{\v{p}_j}(\v{x})\ \forall j\in\{1, ...,n\}\},\ \forall i\in\{1, ...,n\}.
$$

\def\Mah{\mathrm{Mah}}

It follows that all anisotropic Voronoi cells are non-empty as it is the case for $k$-bag Bregman Voronoi diagrams.

Hence, the site weights of a $k$-bag Bregman Voronoi diagram sparsely define a {\em tensor divergence} that indicates  how divergences should be measured locally from the respective bag of divergences. Noteworthy, our study of $k$-bag Bregman Voronoi diagrams shows that the anisotropic Voronoi diagram also admits a {\em second-type} anisotropic Voronoi diagram, induced by the respective dual Legendre functions of the Bregman basis functions of the quadratic distance monomials. 
The Legendre dual of a quadratic distance function $d_{\m{M}}(\v{p},\v{q})=(\v{p}-\v{q})^T\m{M}(\v{p}-\v{q})$ induced by positive-definite matrix $\m{M}$ is the quadratic distance $d_{\m{M}^{-1}}$.
(Matrix $\m{M}$ is itself usually obtained as the inverse of a variance-covariance matrix $\bm\Sigma$ in so-called Mahalanobis distances.)

\section{Bregman triangulations 
\label{sec:BregmanTriangulation}}

\begin{figure}
\centering
\includegraphics[bb=0 0 460 449, width=6cm]{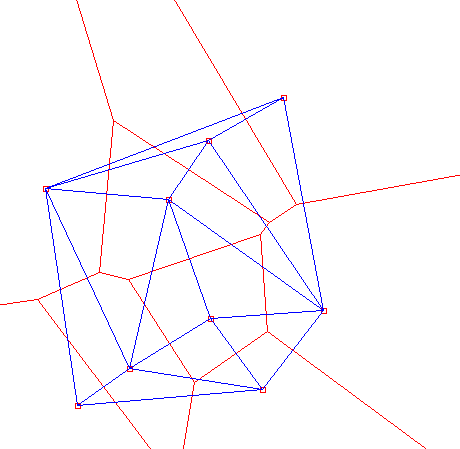}
\caption{Ordinary Voronoi diagram (red) and geometric dual Delaunay triangulation (blue). \label{fig:vordel}}
\end{figure}

Consider the Euclidean Voronoi diagram $\vor (\set{S})$ of a finite set
$\set{S}$ of points of $\R ^d$ (called sites). Let $f$ be a face of
$\vor (\set{S})$ that is the intersection of $k$ $d$-cells of $\vor
(\set{S})$.  We associate to $f$ a dual face $f^*$, namely the convex
hull of the sites associated to the subset of cells. If no subset of
$d+2$ sites lie on a same sphere, the set of dual faces (of dimensions
0 to $d$) constitutes a triangulation embedded in $\R ^d$ whose
vertices are the sites. This triangulation is called the {\em Delaunay
triangulation} of $\set{S}$, noted $\del (\set{S})$.  The
correspondence defined above between the faces of $\vor (\set{S})$ and
those of $\del (\set{S})$ is a bijection that satisfies: $f\subset g
\Rightarrow g^* \subset f^*$.
We say that $\del (\set{S})$ is the {\em geometric dual} of $\vor
(\set{S})$. See Figure~\ref{fig:vordel}.

A similar construct is known also for power diagrams.  Consider the
power diagram of a finite set of balls of $\R ^d$. In the same way as
for Euclidean Voronoi diagrams, we can associate a triangulation dual
to the power diagram of the balls.  This triangulation is called the
{\em regular triangulation} of the balls. The vertices of this triangulation
are the centers of the balls whose cell is non empty.

We derive two triangulations from Bregman Voronoi diagrams. One has
straight edges and captures some important properties of the Delaunay
triangulation. However, it is not always the geometric dual of the
corresponding Bregman Voronoi diagram. The other one has curved
(geodesic) edges and is the geometric dual of the Bregman Voronoi
diagram.

\subsection{Bregman Delaunay triangulations}

Let $\hat{\set{S}}$ be the lifted image of $\set{S}$ and let $\set{T}$
be the lower convex hull of $\hat{\set{S}}$, i.e. the collection of
facets of the convex hull of $\hat{\set{S}}$ whose supporting
hyperplanes are below $\hat{\set{S}}$. We assume in this section that
$\mathcal{S}$ is in {\em general position} if there is no subset of $d+2$
points lying on a same Bregman sphere. Equivalently (see Lemma
\ref{lem-lift-sphere}), $\mathcal{S}$ is in { general position} if
no subset of $d+2$ points $\hat{\v{p}}_i$ lying on a same hyperplane.

\begin{figure}
\centering
\includegraphics[bb=0 0 511 511, width=7cm]{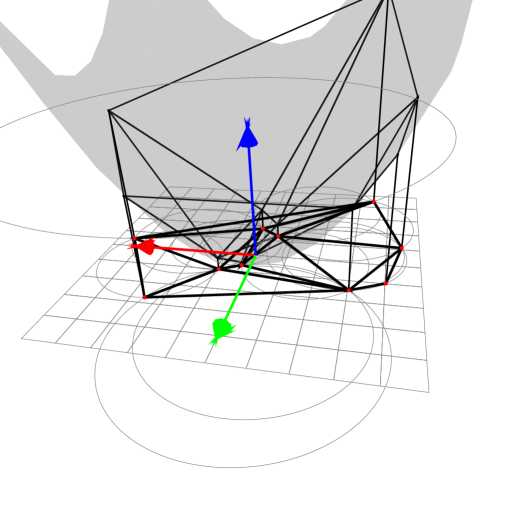}\includegraphics[bb=0 0 511 511, width=7cm]{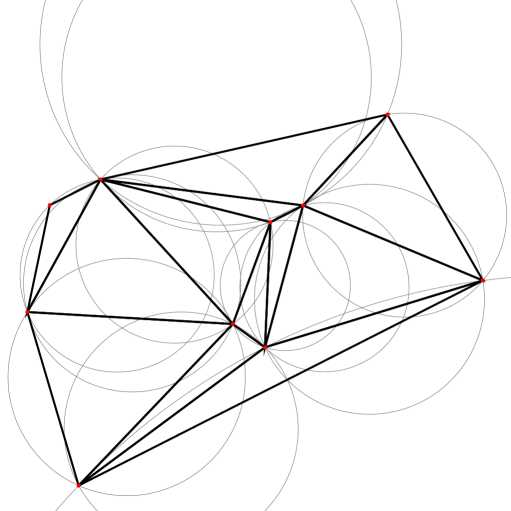}
\caption{Bregman Delaunay triangulation as the projection of the
convex polyhedron $\mathcal{T}$.\label{fig:bvdfrompoly}}
\end{figure}

Under the general position assumption, each vertex of $\mathcal{H}=\cap_{i}
H_{\v{p}_i}^{\uparrow}$ is
the intersection of exactly $d+1$ hyperplanes and the faces of
$\mathcal{T}$ are all {\em simplices}.  Moreover the vertical
projection of $\mathcal{T}$ is a triangulation $\del_F(\mathcal{S})=
\proj_{\set{X}}(\mathcal{T})$ of $\mathcal{S}$ embedded in $\set{X}\subseteq\R^d$. Indeed, since the restriction of $\proj_{\set{X}}$ to $\mathcal{T}$
is bijective, $\del_F(\mathcal{S})$ is a simplicial complex embedded
in $\mathcal{X}$. Moreover, since $F$ is convex, $\del_F(\mathcal{S})$
covers the (Euclidean) convex hull of $\mathcal{S}$, and the set of
vertices of $\mathcal{T}$ consists of all the
$\hat{\v{p}}_i$. Consequently, the set of vertices of
$\del_F(\mathcal{S})$ is $\mathcal{S}$.  We call $\del_F(\mathcal{S})$
the {\em Bregman Delaunay triangulation} of $\mathcal{S}$ (see
Fig. \ref{fig:bvdfrompoly}).  When $F(\vector{x})=||\vector{x}||^2$,
$\del_F(\mathcal{S})$ is the Delaunay triangulation dual to the
Euclidean Voronoi diagram.  This duality property holds for symmetric
Bregman divergences (via polarity) but not for general Bregman divergences.

We say that a Bregman sphere $\sigma$ is {\em empty} if the open ball
bounded by $\sigma$ does not contain any point of $\mathcal{S}$. The
following theorem extends a similar well-known property for Delaunay
triangulations whose proof (see, for example~\cite{compgeom-1998}) can be extended in a straightforward way to 
Bregman triangulations using the lifting map introduced in
Section~\ref{sec:SpherePolarity}.

\begin{theorem}
The first-type Bregman sphere circumscribing any simplex of $\del_F (\mathcal{S})$ is empty. 
$\del_F (\mathcal{S})$ is the only triangulation of $\mathcal{S}$ with this property  when $\mathcal{S}$ is in general position.
\end{theorem}

Several other properties of Delaunay
triangulations extend to Bregman triangulations. We list some of them.

\begin{theorem}[Empty ball]
Let $\nu$ be a subset of at most $d+1$ indices in $\{ 1,\ldots
,n\}$.  The convex hull of the associated points $\v{p}_i$, $i\in
\nu$, is a simplex of the Bregman triangulation of $\mathcal{S}$ 
iff there exists an
empty Bregman sphere $\sigma$ passing through the $\v{p}_i$, $i\in
\nu$.
\end{theorem}

The next property exhibits a local characterization of Bregman
triangulations.  Let $T(\mathcal{S})$ be a triangulation of
$\mathcal{S}$. We say that a pair of adjacent facets $f_1=(f,\v{p}_1)$
and $f_2= (f, \v{p}_2)$ of $T(\mathcal{S})$ is regular iff $\v{p}_1$
does not belong to the open Bregman ball circumscribing $f_2$ and
$\v{p}_2$ does not belong to the open Bregman ball circumscribing
$f_1$ (the two statements are equivalent
for symmetric Bregman divergences).

\begin{theorem}[Locality]
Any triangulation of a given set of points $\mathcal{S}$ (in general position) whose pairs of
facets are all regular is the Bregman triangulation of $\mathcal{S}$.
\end{theorem}

Let $\mathcal{S}$ be a given set of points, $\del_F(\mathcal{S})$ its Bregman triangulation,
and $\mathcal{T}(\mathcal{S})$ the set of all   triangulations of $\mathcal{S}$. We
define the Bregman radius of a $d$-simplex $\tau$ as the radius noted
$r(\tau )$ of the smallest Bregman ball containing $\tau$. The
following result is an extension of a result due to Rajan for Delaunay
triangulations \cite{DBLP:journals/dcg/Rajan94}.

\begin{theorem}[Optimality]
We have $\del_F(\mathcal{S}) = \min_{T\in \mathcal{T}(\mathcal{S})} \max_{\tau\in T} r(\tau )$.
\end{theorem}

The proof mimics Rajan's proof~\cite{DBLP:journals/dcg/Rajan94} for the case of Delaunay triangulations.

\subsection{Bregman geodesic triangulations}

\begin{figure}
\centering
\begin{tabular}{cc}
\fbox{\includegraphics[bb=0 0 470 463, width=6cm]{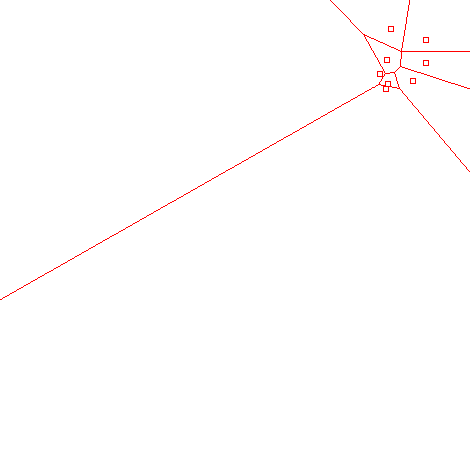}} &
\fbox{\includegraphics[bb=0 0 467 456, width=6cm]{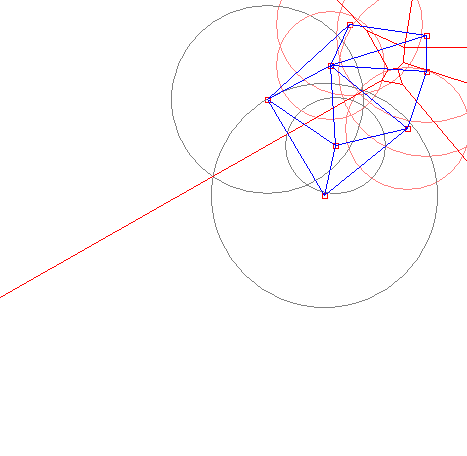}}\\
(a) & (b)
\end{tabular}
\caption{First-type Kullback-Leibler Bregman Voronoi diagram (a) obtained from the corresponding power diagram (b), and its associated dual regular triangulation rooted at gradient vertices (blue).\label{fig:powerdual}}
\end{figure}

\begin{figure}
\centering
\begin{tabular}{ccc}
\includegraphics[bb=0 0 511 511 , width=5cm]{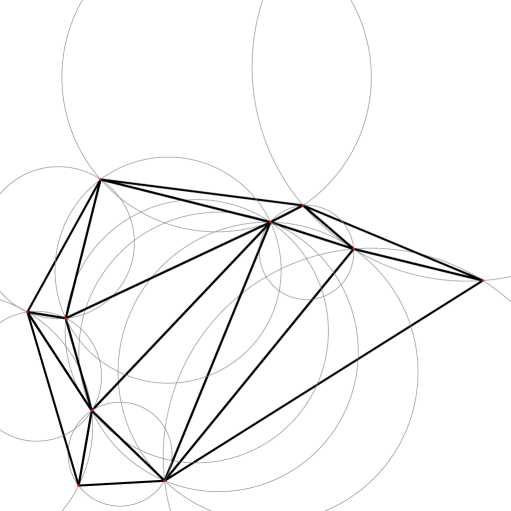} & 
\includegraphics[bb=0 0 511 511 ,width=5cm]{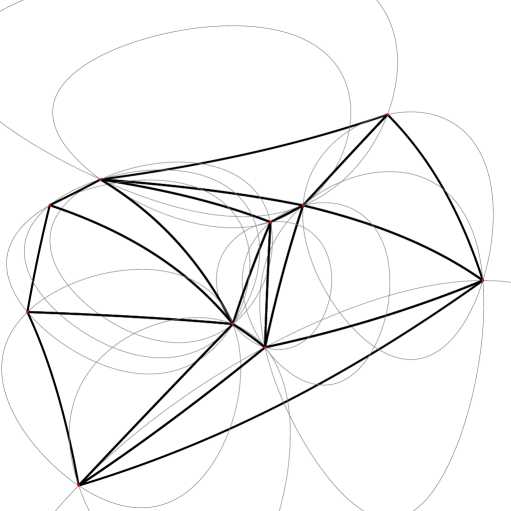} & 
\includegraphics[bb=0 0 511 511 ,width=5cm]{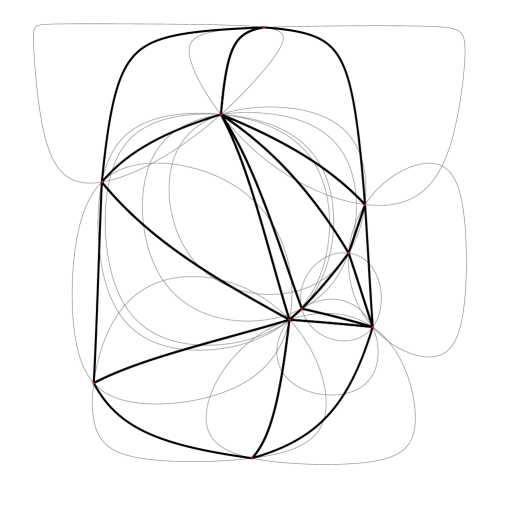}\\
(a) Ordinary Delaunay & (b) Exponential loss & (c) Hellinger-like divergence
\end{tabular}
\caption{An ordinary Delaunay triangulation~(a) and two Bregman geodesic
triangulations for the exponential loss~(b) and for the Hellinger-like 
divergence~(c). }
\label{fig:delaunayl2exp}
\end{figure}

We have seen in Section~\ref{sec:PowerDiagram} that the Bregman
Voronoi of a set of points $\set{S}$ is the power diagram of a set of
balls $\set{B}'$ centered at the points of $\set{S}'$
(Theorem~\ref{th-power}). Write $\reg_F (\set{B}')$ for the dual
regular triangulation dual to this power diagram. This triangulation\footnote{Applet at \url{http://www.csl.sony.co.jp/person/nielsen/BVDapplet/}}
is embedded in $\set{X}'$ and has the points of $\set{S}'$ as its
vertices (see Figure~\ref{fig:powerdual}).  The image of this triangulation by $\v{\nabla}^{-1}F$ is a
{\it curved triangulation} whose vertices are the points of
$\set{S}$. The edges of this curved triangulation are geodesic arcs
joining two sites (see Section~\ref{sec:BVD:geodesicbisector}). We
call it the {\em Bregman geodesic triangulation} of $\set{S}$, noted
$\del'_F(\set{S})$ (see Figure~\ref{fig:delaunayl2exp}).

\begin{theorem}
The Bregman geodesic triangulation $\del'_F(\set{S})$ is the geometric
dual of the 1st-type Bregman Voronoi diagram of $\set{S}$.
\end{theorem}

\begin{proof}
We have, noting  $\stackrel{*}{\equiv}$ for the dual mapping, and using 
Theorem~\ref{th-power}
\[ \vor_F(\set{S}) \equiv \pow (\set{B}') \stackrel{*}{\equiv}
\reg (\set{B}') = \gradF (\del'_F (\set{S})) .\]
\end{proof}

Observe that $\del'_F(\set{S})$ is, in general, distinct from
$\del_F(\set{S})$, the Bregman Delaunay triangulation introduced in
the previous section. However, when the divergence is symmetric, both
triangulations are combinatorially equivalent and dual to the Bregman
Voronoi diagram of $\set{S}$. Moreover, they coincide exactly when $F$ is the
squared Euclidean distance.

\section{Applications\label{sec:applications}}

In this section, we give some applications related to computational geometry and machine learning.

\subsection{Centroidal Bregman Voronoi diagrams and Lloyd quantization\label{sec:centroidal}}

Let $\mathcal{D}$ be a domain of $\mathcal{X}$ and $p(\xx)$ be a
density function defined over $\mathcal{D}$.  We define the {\em
 Bregman centroid} of $\mathcal{D}$ as the point $\v{c}^*\in
\mathcal{D}$ such that $\v{c}^*= \argmin_{\v{c}\in \mathcal{D}}\;
\int_{\v{x}\in\mathcal{D}} p(\xx) D_F (\v{x}||\v{c})\, \dvx$.  The
following lemma states that the mass Bregman centroid of
$\mathcal{D}$ is {\em uniquely} defined and {\em independent} of $F$.

\begin{lemma}
The Bregman centroid of $\mathcal{D}$
coincides with the mass centroid of $\mathcal{D}$.
\end{lemma}

\begin{proof}
\begin{eqnarray*}
\nabla_{\v{c}} \int_{\v{x}\in\mathcal{D}} p(\xx)\;D_F (\v{x}||\v{c})\, \dvx  
&=& \nabla_{\v{c}} \int_{\v{x}\in\mathcal{D}}p(\xx)\;( F(\v{x})-F(\v{c})-
                \dotproduct{\v{x}-\v{c}}{\gradF (\v{c})} ) \dvx\\
&=& - \int_{\v{x}\in\mathcal{D}} p(\xx)\;\hessF (\v{c}) (\v{x}-\v{c})\dvx\\ 
&=& - \hessF (\v{c})( \int_{\v{x}\in\mathcal{D}}  p(\xx)\;\v{x}\dvx  - \v{c} \int_{\v{x}\in\mathcal{D}} p(\xx)\;\dvx ).
\end{eqnarray*}

Hence, $\v{c}^*=\frac{ \int_{\v{x}\in\mathcal{D}}  p(\xx)\;\v{x}\dvx}{\int_{\v{x}\in\mathcal{D}}  p(\xx)\;\dvx}$.
\end{proof}

When $\xx$ is a random variable following the probability density
$p(\xx)$, $\int_{\v{x}\in\mathcal{D}} p(\xx)\;D_F (\v{x}||\v{c})\, \dvx$
is called the {\em distortion rate} associated to the representative
$\v{c}$, the optimal distortion-rate function
$\int_{\v{x}\in\mathcal{D}} p(\xx)\;D_F (\v{x}||\v{c}^*)\, \dvx$ is
called the {\em Bregman information}, and $\v{c}^*$ is called the {\em
Bregman representative}.  The above result states that the optimal
distortion rate exists and does not depend on the choice of the Bregman
divergence, and that the Bregman representative $\v{c}^*$ is the
expectation $E(\xx)$ of $\xx$.  This result extends an analogous
result in the discrete case (finite point sets) studied
in~\cite{j-cbd-2005}.

Computing a centroidal Bregman Voronoi diagram of $k$ points can be
done by means of Lloyd's algorithm~\cite{kmeans-1982}. We select an
initial set of $k$ points. Then, we iteratively compute a Bregman
Voronoi diagram and move the sites to the Bregman centroids of the
corresponding cells in the diagram. Upon convergence, the output of
the algorithm is a local minimizer of $f((\v{p}_i,V_i), i=1,\ldots ,k)
= \sum_{i=1}^{k} \int_{\v{x}\in V_i} D_F (\v{x}||\v{p}_i)\, \dvx$ ,
where $\{ \v{p}_i\}_{i=1}^k$ denotes any set of $k$ points of
$\mathcal{X}$ and $\{ V_i\}_{i=1}^k$ denotes any tesselation of
$\mathcal{X}$ into $k$ regions. See \cite{dfg-cvt-99} for a further
discussion and applications of centroidal Voronoi diagrams.

\subsection{$\varepsilon$-nets}

Lloyd's algorithm intends to find a best set of $k$ points for a given
$k$ so as to minimize a least-square criterion. Differently, we may
want to sample a compact domain $\mathcal{D}\subset \set{X}$ up to a given precision
while minimizing the number of samples. Instead of a least-square
criterion, we define the error associated to a sample $P$ as ${\rm
error} (P) =
\max_{\v{x}\in \mathcal{D}}\, \min_{\v{p}_i\in P}
D_F(\v{x}||\v{p}_i)$.  A finite set of points $P$ of $\mathcal{D}$ is
an {\em $\varepsilon$-sample} of $\mathcal{D}$ iff ${\rm error} (P) \leq
\varepsilon$.  

An $\varepsilon$-sample $P$ is called an {\em $\varepsilon$-net} if it
satisfies the sparsity condition: $\max (D_F(\pp || \qq ), D_F(\qq
||\pp )) >\varepsilon$ for any two points $\pp$ and $\qq$ in $P$.

We will see how to construct an $\varepsilon$-net.
For simplicity, we assume in the rest of the section that
$\mathcal{D}$ is a convex polytope.  Extending the results to more
general domains is possible.

Let $P\subset \mathcal{D}$, $\vor_F (P)$ be the Bregman Voronoi diagram
of $P$ and $\vor_{F|{\mathcal{D}}}(P)$ be its restriction to $\mathcal{D}$.
Write $V$ for the set of vertices of 
$\vor_{F|{\mathcal{D}}}(P)$. $V$ consists of vertices of $\vor_F (P)$ and
intersection points between the edges of $\vor_F (P)$ and the boundary of
$\mathcal{D}$. The following lemma states that ${\rm error} (P)$ can
be computed by examining only a finite number of points, namely the
points of $V$.

\begin{lemma}
\label{th-loose-e-sample}
$ {\rm error} (P) = \max_{\v{v}\in V}\, \min_{\v{p}_i\in P} 
D_F(\v{x}||\v{p}_i)$.
\end{lemma}
\begin{proof}
Let $\v{x}\in \mathcal{D}$, $\v{p}_{\v{x}}$ the point of $P$ closest
to $\v{x}$ and $V_{\v{x}}$ the associated cell of
$\vor_{F|{\mathcal{D}}}(P)$ (which contains $\v{x}$).  $V_{\v{x}}$ is a
bounded polytope whose vertices belongs to $V$. Let $\v{w}$ be the
vertex of $V_{\v{x}}$ most distant from $\v{p}_{\v{x}}$.  We have
$D_F(\v{x}||\v{p}_{\v{x}}) \leq D_F(\v{w}||\v{p}_{\v{x}})$. This is a
consequence of the convexity of $F$ and of the fact that
$D_F(\v{x}||\v{p})$ is measured by the vertical distance between
$\hat{\xx}$ and $H_{\v{p}}$ (Lemma~\ref{lem-code-BD}).
\end{proof}

An $\varepsilon$-net of $\mathcal{D}$ can be constructed by
the following greedy algorithm originally proposed by
Ruppert in the context of mesh generation
\cite{r-draq2d-95}. See also \cite{elpz-fpsis-97}. We initialize the 
sample set $P_0$ with $d$ points of $\mathcal{D}$ lying at distance
greater than $\varepsilon$ from one another. Then, at each step, the
algorithm looks for the point $\v{v}_i$ of $\mathcal{D}$ that is the
furthest (for the considered Bregman divergence) from the current set of samples
$P_i$.  By Lemma \ref{th-loose-e-sample}, this step reduces to looking
at the vertices of $\vor_{F|{\mathcal{D}}}(P_i)$. If
$D_F(\v{x}||\v{v}_i)\leq
\varepsilon$, the algorithm stops.  Otherwise, we take $\v{v}_i$ as a
new sample point, i.e.  $\v{p}_{i+1}=\v{v}_i$, we update the set of
sample points, i.e. $P_{i+1}=P_i\cup
\{ \v{p}_{i+1}\}$, and insert $\v{p}_{i+1}$ in 
the Bregman Voronoi diagram of the sample points.  Upon termination,
the set of sample points $P_t$ satisfies the hypothesis of Lemma
\ref{th-loose-e-sample} and therefore $P_t$ is an $\varepsilon$-sample
of $\mathcal{D}$. Moreover, for any two points $\pp$ and $\qq$ of
$P_t$, we have $D_F(\pp ||\qq)>\varepsilon$ or $D_F(\qq
||\pp)>\varepsilon$, depending on whether $\pp$ has been inserted after
or before $\qq$. Indeed, we only insert a point if its divergence to the
points of the current sample is greater than $\varepsilon$.
Hence, $P_t$ is an $\varepsilon$-net of $\mathcal{D}$.

To prove that the algorithm terminates, we need the following lemma.
Given a Bregman ball $B(\v{c},r)$, we define the biggest Euclidean
ball $EB(\v{c},r')$ contained in $B(\v{c},r)$
and the smallest
Euclidean ball $EB(\v{c},r'')$ containing $B(\v{c},r)$.

\begin{lemma}\label{lemma:fat}
Let $F$ be a strictly convex function of class $C^2$, there are
constants $\gamma '$ and $\gamma ''$ (that do not depend on $\v{c}$
nor on $r$) such that $r'^2\geq \gamma ' r$ and $r''^2\leq \gamma
''r$.
\end{lemma}

\begin{proof}
According to Taylor's formula, there exists a point  $\v{t}$ of the open segment $\v{x}\v{c}$ such that
\[ F(\xx)= F(\v{c})+ \dotproduct{\v{x}-\v{c}}{\gradF (\v{c})} 
+  \frac{1}{2}\; 
(\v{x}-\v{c})^T \hessF(\vector{t}) (\v{x}-\v{c}).\]
Hence,
\begin{equation}
 D_F(\v{x}|| \v{c}) = F(\v{x})-F(\v{c}) - \dotproduct{\v{x}-\v{c}}{\v{c}'} = \frac{1}{2}\; 
(\v{x}-\v{c})^T \hessF(\vector{t}) (\v{x}-\v{c}),
\label{eq-fatness}
\end{equation}
where $\v{t}$ is a point of the open segment $\v{x}\v{c}$.

Since $F$ is strictly convex, the Hessian matrix is positive definite (i.e., $\v{x}^T\hessF(\vector{t})\v{x}>0$ for all $\v{x}$ in $\set{X}$), and the domain
$\mathcal{D}$ being compact, there exist two constants $\eta '$ and
$\eta ''$ such that, for any $\v{y}\in \mathcal{D}$, $0 < \eta '' \leq
||\hessF(\vector{y})||\leq \eta ' $.
If $\| \v{x}-\v{c}\| ^2 > \frac{2r}{\eta ''}$ (Fr\"obenius matrix norm), we deduce from Equation
(\ref{eq-fatness}) that $D_F(\v{x}|| \v{c}) > r$. Therefore,
$B(\v{c},r) \subset EB(\v{c}, \sqrt{\frac{2r}{\eta ''}})$.

If $\| \v{x}-\v{c}\| ^2 \leq \frac{2r}{\eta '}$, we have
using again Equation~(\ref{eq-fatness})
\[ D_F(\v{x}|| \v{c}) \leq \frac{\eta '}{2}\; 
\| \v{x}-\v{c}\| ^2  \leq r.\]
Therefore,
$EB(\v{c}, \sqrt{\frac{2r}{\eta '}}) \subset B(\v{c},r) $.
\end{proof}

Let $\pp$ and $\qq$ be two points such that $D_F(\pp ||\qq)
=r$. Observing that $EB(\pp, r')\subseteq EB(\pp, \| \pp
-\qq\|)\subseteq EB(\pp, r'')$, we  deduce from the above lemma
that
\begin{equation}
\sqrt{\gamma '\, r}\leq \| \pp -\qq\| \leq \sqrt{\gamma ''\, r}
\label{eq-pq-r}
\end{equation}

and
\[ \frac{\gamma '}{\gamma ''} D_F(\pp || \qq )  \leq  D_F(\qq || \pp ) \leq  \frac{\gamma ''}{\gamma '} D_F(\pp || \qq ).\]

Another consequence of the lemma is that the volume of any Bregman ball 
of radius at least $r>0$, is bounded away from 0  (when $F$ is
of class $C^2$). Hence, since $\mathcal{D}$ is compact, the algorithm
cannot insert infinitely many points and therefore terminates. Moreover, 
the size of the sample
output by the algorithm can be bounded, as stated in the next lemma.
Write $\mathcal{D}^{\leq\varepsilon}=\{
\v{x}|\  \exists \v{y}\in \mathcal{D}, \| \v{x}-\v{y}\|  \leq
\varepsilon \}$.

\begin{lemma}
If $F$ is of class $C^2$, the algorithm terminates. If $P_t$ denotes
the final set of sample points, we have $|P_t| = O\left(\frac{{\rm vol} (\mathcal{D})}{\epsilon ^{d/2}}\right)$.
\end{lemma}

\begin{proof}
We have already shown that the algorithm terminates. Let $P_t$ be the set of points that have been inserted by the algorithm, excluding the initial set (of constant size).
Let $\tau (\xx) = \inf \{r : |EB(\xx,r)\bigcap P_t| \geq 2\} $ and $B_p=EB(\pp,
\frac{\tau (\pp)}{2})$, $\pp\in P_t$. It is easy to see that $\tau$ is
1-Lipschitz and that the Euclidean balls $B_p$, $\pp\in P_t$ are
disjoint. 
Let $\qq$ be a point of $P_t$ closest to $\pp$~: $\tau (\pp)= \| \pp-\qq\|$ and,
as noticed above, $\max (D_F( \pp||\qq), D_F(\qq||\pp)) >
\varepsilon$. Eq.~\ref{eq-pq-r} then implies that 
$\tau(\pp)=\| \pp -\qq\| \geq \sqrt{\gamma '\, \varepsilon
}$. Consider now the midpoint $\v{m}$ of $\pp\qq$ and write $\v{t}$
for the point of $P_t$ that minimizes $D_F(\v{m}||.)$. Since
$\mathcal{D}$ is convex, $\v{m}\in \mathcal{D}$ and, according to the
definition of $\qq$, $\| \v{m}-\pp\| \leq \|
\v{m}-\v{t}\|$. Eq.~\ref{eq-pq-r} and the fact that $P_t$ is an
$\varepsilon$-sample of $\mathcal{D}$ then yield $ \| \v{m}-\v{t}\|
\leq \sqrt{\gamma ''\,
\varepsilon}$. In summary, we have

\begin{equation}
\sqrt{\gamma '\, \varepsilon }\leq \tau(\pp)=\| \pp -\qq\| \leq 2\sqrt{\gamma ''\, \varepsilon}.
\label{eq-tau}
\end{equation}
The right inequality shows that all the balls $B_{\v{p}}$, $\v{p}\in P_t$, are
contained in $\mathcal{D}^{\leq\eta}$ where $\eta =\sqrt{\gamma ''\, \varepsilon}$.
We can now bound the size of $P_t$.
\begin{tabbing}
\mbox{}\hspace{2cm} $\int _{\mathcal{D}^{\leq\eta}} \frac{d\xx}{\tau ^{d}(\xx)}$ \=  $\geq 
\sum _{\pp\in P_t}\int _{ B_p\cap \mathcal{D}^{\leq\eta}}
\frac{d\xx}{\tau ^{d}(\xx)} $ \hspace{5mm}{\small (the balls $B_p$ have  disjoint interiors)}\\
\mbox{} \\
\> $ \geq   \sum _{\pp\in P  } \frac{{\rm vol} (B_p\cap \mathcal{D}^{\leq \eta})}{(\frac{3}{2}\tau(\pp))^{d}}$
\hspace{5mm} {\small ($\tau (\xx) \leq \tau (\pp) + \| \pp-\xx\| \leq \frac{3}{2}\; \tau (\pp)$)}\\
\mbox{} \\
\> $\geq \frac{C}{3^d} \; |P_t|$ 
\end{tabbing}
where $C = \frac{\pi^p}{p!}$ if $d=2p$ and $C= \frac{2^{2p-1}(p-1)!\,
\pi^{p-1}}{(2p-1)!}$ if $d=2p-1$.

Using again the Lipschitz property of $\tau$ and  Eq~\ref{eq-tau},   we
have for all $\xx\in B_p$
$$ \tau (\xx)\geq \tau (\pp) - \| \xx-\pp\| \geq \frac{1}{2}\; \tau (\pp)  \geq \frac{1}{2}\, \sqrt{\gamma '\,\varepsilon}
$$

We deduce
\[ |P_t| \leq \left( \frac{6}{\sqrt{\gamma '}}\right) ^{d}\;\frac{1}{C\varepsilon ^{d/2}}\; {\int}_{\mathcal{D}^{\leq \eta}}  \;\;d\xx . \]
\end{proof}

\def\vol{\mathrm{Vol}}

A geometric  object $O$ is said $\alpha$-fat~\cite{realistic-2002} if the ratio $\frac{r^+}{r^-}$ of the radius $r^+$ of the smallest  ball enclosing $O$ over the radius $r^-$ of the largest  ball inscribed in $O$ is bounded by $\alpha$:  $\frac{r^+}{r^-}\leq \alpha$. Euclidean balls are therefore $1$-fat, namely the fattest objects.
It has been shown that considering the fatness factor for a set of objects yields in practice efficient tailored data-sensitive algorithms~\cite{realistic-2002} by avoiding bad configurations of  sets of skinny objects.
A direct consequence of Lemma~\ref{lemma:fat} is that Bregman balls (in fixed dimensions) are fat (i.e., $\alpha=O(1)$) on any {\it compact} domain:

\begin{corollary}
For $C^2$ Bregman generator functions,  Bregman balls  on any compact domain are fat. 
\end{corollary}

\begin{proof}
Indeed, consider any Bregman ball defined on a compact domain for a $C^2$ strictly convex and differentiable Bregman generator function $F$.
Its fatness $\alpha$ is upper bounded by $\sqrt{\frac{\gamma'}{\gamma''}}$, where $\gamma'$ and $\gamma''$ are the two constants (depending on $F$ and $\set{D}$) of Lemma~\ref{lemma:fat}. 
Recall that Lemma~\ref{lemma:fat} considers concentric Euclidean balls ham sandwiching a Bregman ball, all centered at position $\v{c}$. 
We have $\alpha\leq\frac{r^+}{r^-}\leq \frac{r^+}{r_{\v{c}}^-} \leq \frac{r_{\v{c}}^+}{r_{\v{c}}^-}=O(1)$ since $r_{\v{c}}^-\leq r^-$ and $r_{\v{c}}^+\geq r^+$, where $r_{\v{c}}^+$ (respectively, $r_{\v{c}}^-$) denote  the radius of the smallest enclosing (respectively, largest inscribed) Euclidean ball centered at $\v{c}$. 
The fatness property simply means that we can cover any Bregman ball by a constant number of (convex) Euclidean balls. 
\end{proof}

Thus, since Bregman balls are fat on compact domains, we can build efficient data-structures for point location with applications to piercing (geometric $0$-transversal) and others, as described in~\cite{ekns-ddsfo:2000}.

\subsection{VC-dimension, classification and learning}

Some important classification rules rely on Voronoi diagrams; furthermore, the analysis of
classification rules (complexity or statistical generalization) sometimes makes use
of concepts closely related to Voronoi diagrams. Extending the rules and analyses to arbitrary Bregman
divergences, with important related consequences (such as the eventual lost of convexity)
is thus particularly interesting for classification, and we review here some notable consequences.

In supervised classification, we are generally interested in capturing the joint structure of ${\set{X}}$
and a set of \textit{classes}, $\{0, 1\}$ in the simplest case. For this objective, we build
\textit{representations of concepts}, \textit{i.e.} functions that map ${\set{X}}$ to
the set of classes. A concept class ${\set{H}}$ is a set of concept representations
$h : {\set{X}} \rightarrow \{0, 1\}$; for example, should $h$ be a Bregman ball, it would classify 0 the points 
outside the ball, and 1 the points inside. Armed with these definitions, our supervised classification problem
becomes the following one. A so-called \textit{target} concept, $c$, which is unknown, labels the
points of ${\set{X}}$; we have access to its labeling throughout a sampling process: we
retrieve \textit{examples} ({\it i.e.}, pairs $(\bm{x}, c(\bm{x}))$), independently at random, according to 
some {\em unknown} but {\em fixed} distribution ${\set{D}}$ over the set $\{(\bm{x},c(\bm{x})) : \bm{x} \in {\set{X}}\}$.
The question is: what are the conditions on ${\set{H}}$ that 
guarantee the possibility to build, within reasonable time, some $h \in {\set{H}}$ agreeing as best as possible with $c$, with
high probability? While the complexity requirement is usual in computer science,
the fact that we require adequacy with high probability better than systematically is also a necessary requirement,
as there is always the possibility of an extremely bad sampling that would prevent any efficient learning 
(\textit{e.g.} we have drawn the same example all the time). In general, rather than directly sampling the domain, we 
work with a finite data set ${\set{S}}$ of examples which is supposed to be sampled this way. 

From the statistical standpoint, learning requires to find a good balance between the accuracy, \textit{i.e.}
the goodness-of-fit of $h$ as measured on ${\set{S}}$, and the \textit{capacity} of ${\set{H}}$, 
\textit{i.e.} its ability to \textit{learn} (or fit in generalization) the data with the smallest number 
of errors. Consider for example geometric figures in the plane and the ``square'' concept. 
Intuitively, an ${\set{H}}$ with too large capacity is like the person who picks a huge quantity of geometric figures including squares,
memorizes each of them, and then rejects every square that would not exactly be in its collection (edge lengths, colors, etc.). 
An ${\set{H}}$ with
too little capacity is like the lazy person who keeps as sole concept the fact that squares have four edges. Both extremal situations
mean little generalization capabilities, but for different reasons. 

There have been intensive lines of works on the measures of this
capacity, and one of the most popular is the VC-dimension
\cite{dglAP}.  Informally, the VC-dimension of ${\set{H}}$ is the size
of the largest dataset $\set{S}$ for which ${\mathcal{H}}$ {\em
shatters} $\set{S}$, i.e.  for which ${\mathcal{H}}$ contains all the
classifiers that could perform any of the $2^{|\set{S}|}$ possible
labelings of the data. To be more formal, let
$\Pi_{{\set{H}}}({\set{S}}) =
\{(h(\bm{p}_1), h(\bm{p}_2), ..., h(\bm{p}_n))\ |\ h \in {\set{H}}\}$
denote the set of all distinct tuples of labels on ${\set{S}}$ that
can be performed by elements of ${\mathcal{H}}$.  While it always
holds that $\Pi_{{\set{H}}}(n) \leq 2^n$, the maximal $n$ for which
$\Pi_{{\set{H}}}(n) = 2^n$ is the VC-dimension of ${\set{H}}$,
$\textsc{VCdim}({\set{H}})$. The importance of the VC-dimension comes
from the fact that it allows to bound the behavior of the empirical
optimal classifier in a distribution-free manner \cite{dglAP}.  In
particular, if the VC-dimension is finite, the average error
probability of the empirical optimal classifier tends to 0 when the
size of the training data set increases. The following lemma proves
that the VC-dimension of Bregman balls is the same as for linear separators,
and this does not depend on the choice of $F$.

\begin{theorem}
The VC dimension of the class of all Bregman balls $B_F$ of $\R ^d$
(for any given strictly convex and differentiable function $F$) is $d+1$.  
\label{th-VCdim}
\end{theorem}

\begin{proof}
We use the lifting map introduced in
Section~\ref{sec:SpherePolarity}. Given a set $\set{S}$ of points in
$\R ^d$, we lift them onto $\mathcal{F}$, obtaining $\hat\set{S}\in \R ^{d+1}$.

Let $B_F$ be a Bregman ball and write $\sigma$ for the Bregman sphere
bounding $B_F$. From Lemma~\ref{lem-lift-sphere}, we know that, for
any $\pp\in \R^d$, $\pp\in B$ iff $\hat{\pp}\in
H_{\sigma}^{\downarrow}$.  For a given function $F$, let
$\mathcal{B}_F$ denote the set of all Bregman balls, and let
$\mathcal{H}_F$ denote the set of all lower halfspaces of $\R^{d+1}$.
It follows from the observation above that $\mathcal{B}$ shatters
$\set{S}$ iff $\mathcal{H}$ shatters $\hat\set{S}$.  Hence the VC
dimension of $\mathcal{B}$ over the sets of points of $\R^d$ is equal
to the VC dimension of $\mathcal{H}$ over the sets of points of
$\mathcal{F}\subset \R^{d+1}$.

Since the points of $\hat\set{S}$ are in convex position, they are
shattered by $\mathcal{H}$ iff the affine hull of their convex hull is
of dimension strictly less than the dimension of the embedding space,
i.e. $d+1$, which happens iff $|\set{S}| < d+2$.  Indeed otherwise,
the subset of vertices of any facet of the upper convex hull of
$\hat\set{S}$ cannot be obtained by intersecting $\hat\set{S}$ with a
lower halfspace (an upper halfspace would be required). Hence, the VC
dimension of Bregman balls is at most $d+1$.  

It is exactly $d+1$ since any set of $d+1$ points on $\mathcal{F}$ in
general position generates a $d$-dimensional affine hull $\set{A}$
that cannot be shattered by less than $d+1$ hyperplanes of
$\set{A}$. The same result plainly holds for hyperplanes of $\R
^{d+1}$ since we can associate to each hyperplane $h$ of $\set{A}$
a hyperplane $H$ of $\R ^{d+1}$ such that $h=H\cap\set{A}$. 
\end{proof}

This result does not fall into the general family of VC bounds for concept classes parameterized by polynomial-based predicates \cite{gjBT}, it is mostly exact, and it happens not to depend on the choice of the Bregman divergence. This has a direct consequence for classification, which is all the more important as Bregman balls are not necessarily convex (see Figure~\ref{fig:convexballs}). Because the capacity of Bregman balls is not affected by the divergence, if we fit this divergence in order to minimize the empirical risk (risk estimated on ${\set{S}}$), then there is an efficient minimization of the true risk (risk estimated on the full domain ${\set{X}}$), as well. There is thus little impact (if any) on overfitting, one important pitfall for classification, usually caused by over-capacitating the classifiers by tuning too many parameters.

Some applications of our results in supervised learning also meet one of the oldest classification rule: the $k$-Nearest Neighbors ($k$-NN) rule \cite{fhDA}, in which a new observation receives the majority class among the set of its $k$ nearest neighbors, using \textit{e.g.} $k$-order Voronoi diagrams of ${\mathcal{S}}$ (Section~\ref{sec:generalizedBVD}). Various results establish upperbounds for the $k$-NN rule that depend on the Bayes risk (the true risk of the best possible rule)~\cite{dglAP}. The choice of the proximity notion between observations (it is often not a metric for complex domains) is crucial: if it is too simple or oversimplified, it degrades the $k$-NN results and may even degrade Bayes risk as well; if it is too complicated or complexified, it may degrade the test results via the capacity of the rule. Searching for accurate ``distance'' notions has been an active field of research in machine learning in the past decade \cite{wmIH}. Our results on the linearity of the Bregman Voronoi diagrams  essentially show that we can mix arbitrary Bregman divergences for heterogenous data (mixing binary, real, integer values, etc.) without losing anything from the capacity standpoint.\\

Range spaces of finite VC-dimensions have found numerous applications
in Combinatorial and Computational Geometry. We refer to Chazelle's
book for an introduction to the subject and references
wherein~\cite{Cha00}. In particular, Br\"onnimann and
Goodrich~\cite{bg-aoscf-95} have proposed an almost optimal solution
to the disk cover algorithm, i.e.  to find a minimum number of disks
in a given family that cover a given set of
points. Theorem~\ref{th-VCdim} allows to extend this result to
arbitrary Bregman ball cover (see also~\cite{hittingset-2005}).

\section{Conclusion\label{sec:Conclusion}}

We have defined the notion of Bregman Voronoi diagrams and showed how
these geometric structures are a natural extension of ordinary Voronoi
diagrams. Bregman Voronoi diagrams share with their Euclidean analogs
surprisingly similar combinatorial and geometric properties.  We hope
that our results will make Voronoi diagrams and their relatives
applicable in new application areas.  In particular, Bregman Voronoi
diagrams based on various entropic divergences are expected to find
 applications in information retrieval (IR), data mining, knowledge
discovery in databases, image processing
(e.g., see~\cite{cccg98-inaba-geometric}). The study of Bregman Voronoi
diagrams raises the question of revisiting computational geometry
problems in this new light.  This may also allow one to tackle
uncertainty ('noise') in computational geometry for fundamental
problems such as surface reconstruction or pattern matching.

A limitation of Bregman Voronoi diagrams is their combinatorial
complexity that depends exponentially on the dimension. Since many
applications are in high dimensional spaces, building efficient
data-structures is a major avenue for further research.

\section*{Acknowledgements}
Fr\'ed\'eric Chazal, David Cohen-Steiner and Mariette Yvinec are gratefully acknowledged for 
their comments on this paper. The work by the second author has been partially supported by the project GeoTopAl (1555) of the Agence Nationale de la Recherche (ANR).


\begin{thebibliography}{10}

\bibitem{informationgeometry}
S.~Amari and H.~Nagaoka.
\newblock {\em Methods of Information Geometry}.
\newblock Oxford University Press, ISBN 0-8218-0531-2, 2000.


\bibitem{svc}
A. Ben-Hur, D. Horn, H. T. Siegelmann, and V. Vapnik
\newblock Support Vector Clustering. 
\newblock {\em Journal of Machine Learning Research}, (2):125-137, 2001.





\bibitem{powerdiagrams-1987}
F~Aurenhammer.
\newblock Power diagrams: Properties, algorithms and applications.
\newblock {\em SIAM Journal of Computing}, 16(1):78--96, 1987.

\bibitem{affinevoronoi-1987}
F.~Aurenhammer and H.~Imai.
\newblock Geometric relations among voronoi diagrams.
\newblock In {\em 4th Annual Symposium on Theoretical Aspects of Computer
  Sciences (STACS)}, pp. 53--65,  1987. 

\bibitem{ak-vd-00}
F.~Aurenhammer and R.~Klein.
\newblock {V}oronoi {D}iagrams.
\newblock In J.~Sack and G.~Urrutia (Eds), {\em Handbook of Computational
  Geometry, Chapter V}, pp. 201--290. Elsevier Science Publishing, 2000.

\bibitem{j-cbd-2005}
A.~Banerjee, S.~Merugu, I.~S. Dhillon, and J.~Ghosh.
\newblock Clustering with {B}regman divergences.
\newblock {\em Journal of Machine Learning Research (JMLR)}, 6:1705--1749,
  2005.
  
\bibitem{realistic-2002}  
M. de Berg, M. Katz, F. van der Stappen, and J. Vleugels. 
\newblock Realistic input models for geometric algorithms. 
\newblock {\em Algorithmica} 34:81-97, 2002.


\bibitem{prisme-4504i}
J.-D. Boissonnat and M.~Karavelas.
\newblock On the combinatorial complexity of {Euclidean Voronoi} cells and
  convex hulls of $d$-dimensional spheres.
\newblock In {\em Proc. 14th ACM-SIAM Sympos. Discrete Algorithms (SODA)},
  pp. 305--312, 2003.

\bibitem{DBLP:conf/cccg/BoissonnatWY05}
J.-D. Boissonnat, C.~Wormser, and M.~Yvinec.
\newblock Anisotropic diagrams: {L}abelle {S}hewchuk approach revisited.
\newblock In {\em 17th Canadian Conference on Computational Geometry (CCCG)},
  pp. 266--269, 2005.

\bibitem{compgeom-1998}
J.-D. Boissonnat and M.~Yvinec.
\newblock {\em Algorithmic Geometry}.
\newblock Cambridge University Press, New York, NY, USA, 1998.

\bibitem{bwy-cvd-07}
J.-D. Boissonnat, C. Wormser, and M. Yvinec.
\newblock Curved {Voronoi} diagrams.
\newblock In J.-D. Boissonnat and M. Teillaud (Eds) {\em
  Effective Computational Geometry for Curves and Surfaces}, pp. 67--116.
  Springer-Verlag, Mathematics and Visualization, 2007.

\bibitem{Bregman67}
L.~M. Bregman.
\newblock The relaxation method of finding the common point of convex sets and
  its application to the solution of problems in convex programming.
\newblock {\em USSR Computational Mathematics and Mathematical Physics},
  7:200--217, 1967.

\bibitem{bg-aoscf-95}
H.~Br{\"o}nnimann and M.~T. Goodrich.
\newblock Optimal set covers in finite {VC}-dimension.
\newblock {\em Discrete \& Computational Geometry}, 14(4):463--479, 1995.

\bibitem{Chazelle1993}
B.~Chazelle.
\newblock An optimal convex hull algorithm in any fixed dimension.
\newblock {\em Discrete Computational Geometry}, 10:377--409, 1993.

\bibitem{Cha00}
B.~Chazelle.
\newblock {\em The Discrepancy Method}.
\newblock Cambridge University Press, Cambridge, U.K., 2000.

\bibitem{Csiszar91}
I.~Csisz{\'a}r.
\newblock Why least squares and maximum entropy? An axiomatic approach to
  inference for linear inverse problems.
\newblock {\em Ann. Stat.}, 19:2032--2066, 1991.

\bibitem{dglAP}
L.~Devroye, L.~Gy{\"o}rfi, and G.~Lugosi.
\newblock {\em A {P}robabilistic {T}heory of {P}attern {R}ecognition}.
\newblock Springer, 1996.



\bibitem{dfg-cvt-99}
Q. Du, V. Faber, and M. Gunzburger.
\newblock Centroidal Voronoi tesselations: Applications and algorithms.
\newblock {\em SIAM Review}, 41:637--676, 1999.

\bibitem{ekns-ddsfo:2000}
A. Efrat, M.~J. Katz, F. Nielsen, and M. Sharir.
\newblock Dynamic data structures for fat objects and their applications.
\newblock {\em Comput. Geom. Theory Appl.}, 15(4):215--227, 2000.

\bibitem{elpz-fpsis-97}
Y. Eldar, M. Lindenbaum, M. Porat, and Y.~Y. Zeevi.
\newblock The farthest point strategy for progressive image sampling.
\newblock {\em IEEE Trans. on Image Processing}, 6(9):1305--1315, 1997.

\bibitem{hittingset-2005}
G. Even, D. Rawitz, and S. Shahar.
\newblock Hitting sets when the VC-dimension is small.
\newblock {\em Inf. Process. Lett.}, 95(2):358--362, 2005.

\bibitem{fhDA}
E.~Fix and J.~L. Hodges.
\newblock Discrimatory analysis, nonparametric discrimination.
\newblock Technical Report TR-21-49-004, Rept 4, USAF School of Aviation
  Medicine, Randolph Field, TX, 1951.

\bibitem{gjBT}
P.-W. Goldberg and M.~Jerrum.
\newblock Bounding the {Vapnik-Chervonenkis} dimension of concept classes
  parameterized by real numbers.
\newblock {\em Machine Learning}, 18:131--148, 1995.

\bibitem{cccg98-inaba-geometric}
M. Inaba and H. Imai.
\newblock Geometric clustering models for multimedia databases.
\newblock In {\em Proceedings of the 10th Canadian Conference on Computational
  Geometry (CCCG'98)}, 1998.

\bibitem{DBLP:conf/cccg/InabaI00}
M. Inaba and H. Imai.
\newblock Geometric clustering for multiplicative mixtures of distributions in
  exponential families.
\newblock In {\em Proceedings of the 12th Canadian Conference on Computational
  Geometry (CCCG'00)}, 2000.

\bibitem{abstractvoronoidiagrams-1989}
R.~Klein.
\newblock {\em Concrete and Abstract Voronoi Diagrams}, volume 400 of {\em
  Lecture Notes in Computer Science}.
\newblock Springer, 1989.
\newblock ISBN 3-540-52055-4.

\bibitem{anisotropicvor-2003}
F.~Labelle and J.~R. Shewchuk.
\newblock Anisotropic voronoi diagrams and guaranteed-quality anisotropic mesh
  generation.
\newblock In {\em Proc. 19th Symposium on Computational Geometry (SoCG)}, pages
  191--200, New York, NY, USA, 2003. ACM Press.

\bibitem{lafferty}
J. Lafferty.
\newblock Additive models, boosting, and inference for generalized divergences.
\newblock In {\em Proc. 12th Conference on Computational learning theory}, 125-133, 1999.

 
\bibitem{entboost}
D.-D. Le and S. Satoh.
\newblock Ent-Boost: Boosting Using Entropy Measure for Robust Object Detection.
\newblock In {\em Proc. 18th International Conference on Pattern Recognition}, pp. 602-605, 2006.


\bibitem{kmeans-1982}
S.~P. Lloyd.
\newblock Least squares quantization in {PCM}.
\newblock {\em IEEE Transactions on Information Theory}, 28(2):129--136, 1982.

\bibitem{McMullen1971}
P.~McMullen.
\newblock The maximum numbers of faces of a convex polytope.
\newblock {\em J. Combinatorial Theory, Ser. B}, 10:179--184, 1971.

\bibitem{b-n-visualcomputing-2005}
F. Nielsen.
\newblock {\em Visual Computing: Geometry, Graphics, and Vision}.
\newblock Charles River Media/Thomson Delmar Learning, ISBN 1584504277,  2005.

\bibitem{inria-recherche-1620}
M.~Teillaud O.~Devillers, S.~Meiser.
\newblock The space of spheres, a geometric tool to unify duality results on
  voronoi diagrams.
\newblock Technical Report  No.1620, INRIA, 1992.

\bibitem{vdnormal}
K.~Onishi and H.~Imai.
\newblock Voronoi diagram in statistical parametric space by
  {K}ullback-{L}eibler divergence.
\newblock In {\em Proc. 13th Symposium on Computational Geometry (SoCG)}, pages
  463--465, New York, NY, USA, 1997. ACM Press.

\bibitem{VoronoiExpFamily-1997}
K.~Onishi and H.~Imai.
\newblock Voronoi diagrams for an exponential family of probability
  distributions in information geometry.
\newblock In {\em Japan-Korea Joint Workshop on Algorithms and Computation},
  1997.

\bibitem{cgal:pt-tds3-06}
S. Pion and M. Teillaud.
\newblock 3d triangulation data structure.
\newblock In CGAL~Editorial Board, editor, {\em CGAL-3.2 User and Reference
  Manual}. 2006.

\bibitem{DBLP:journals/dcg/Rajan94}
V.~T. Rajan.
\newblock Optimality of the Delaunay triangulation in $\mathbb{R}^{\mbox{d}}$.
\newblock {\em Discrete {\&} Computational Geometry}, 12:189--202, 1994.

\bibitem{ConvexAnalysis-1970}
R.~T. Rockafellar.
\newblock {\em Convex Analysis}.
\newblock Princeton University Press, Princeton, New Jersey, 1970.

\bibitem{r-draq2d-95}
J.~Ruppert.
\newblock A {Delaunay} refinement algorithm for quality 2-dimensional mesh
  generation.
\newblock {\em J. Algorithms}, 18:548--585, 1995.

\bibitem{voronoidivergence-1998}
K.~Sadakane, H.~Imai, K.~Onishi, M.~Inaba, F.~Takeuchi, and K.~Imai.
\newblock Voronoi diagrams by divergences with additive weights.
\newblock In {\em Proc. 14th Symposium on Computational Geometry (SoCG)}, pages
  403--404, New York, NY, USA, 1998. ACM Press.

\bibitem{s-atubl-94}
M.~Sharir.
\newblock Almost tight upper bounds for lower envelopes in higher dimensions.
\newblock {\em Discrete Comput. Geom.}, 12:327--345, 1994.

\bibitem{wmIH}
D.~Randall Wilson and Tony~R. Martinez.
\newblock Improved heterogeneous distance functions.
\newblock {\em Journal of Artificial Intelligence Research}, 1:1--34, 1997.

\end{thebibliography}
\end{document}